\documentclass[11pt]{article}

\usepackage{tabularx}

\usepackage{siunitx} 
\usepackage[]{apacite}
\usepackage{booktabs}
\usepackage{scalerel,amsmath}
\usepackage{amssymb,amsthm}
\usepackage{eurosym}

\makeatletter
\edef\originalbmathcode{%
    \noexpand\mathchardef\noexpand\@tempa\the\mathcode`\(\relax}
\def\resetMathstrut@{%
  \setbox\z@\hbox{%
    \originalbmathcode
    \def\@tempb##1"##2##3{\the\textfont"##3\char"}%
    \expandafter\@tempb\meaning\@tempa \relax
  }
  \ht\Mathstrutbox@\ht\z@ \dp\Mathstrutbox@\dp\z@
}
\makeatother

\def\yyy{%
  \bgroup\uccode`\~\expandafter`\string-%
  \uppercase{\egroup\edef~{\noexpand\text{\char"2212\relax}}}%
  \mathcode\expandafter`\string-"8000 }

\usepackage[hidelinks, hyperfootnotes=false]{hyperref}
\usepackage{bookmark}

\usepackage{dirtytalk}
\usepackage{pdflscape}
\usepackage{verbatim}\allowdisplaybreaks
\usepackage{graphicx}
  \usepackage{epstopdf}

\usepackage[hmargin=1in,vmargin=1in]{geometry}
\usepackage[comma]{natbib}
\usepackage{cancel}
\usepackage{subcaption}
\usepackage{anyfontsize}
\usepackage{setspace}
\usepackage{appendix}
\usepackage[flushleft]{threeparttable}
\usepackage[]{caption}
\usepackage{comment}
\usepackage{enumitem}   
\usepackage{longtable}
\usepackage{multirow}

\usepackage{chngcntr}
\usepackage{etoolbox}
\ifundef{\abstract}{}{\patchcmd{\abstract}%
    {\quotation}{\quotation\noindent\ignorespaces}{}{}}
\usepackage{color, colortbl}
\definecolor{Navy}{rgb}{0,0,.4}
\definecolor{OliveGreen}{rgb}{0,0.4,0}
\definecolor{Maroon}{rgb}{.75,0,0}

\hypersetup{
  colorlinks   = true, 
  urlcolor     = black, 
  linkcolor    = black, 
  citecolor   = Maroon 
}

\usepackage{bm}
\usepackage{nicefrac}
\usepackage{caption}
\usepackage{fancyhdr}
\usepackage{multirow}
\usepackage{cellspace}

\usepackage{titlesec}

\titlespacing*{\section}{0pt}{0.5\baselineskip}{0.1\baselineskip}
\titlespacing*{\subsection}{0pt}{0.4\baselineskip}{0.1\baselineskip}
\titlespacing*{\subsubsection}
  {0pt}{0.2\baselineskip}{0.4\baselineskip}
  
\titleformat{\subsubsection}[runin]
        {\normalfont\bfseries}
        {\thesubsection}
        {0.5em}
        {\hspace{2em}}
        []

\newcommand\cites[1]{\citeauthor{#1}'s\ (\citeyear{#1})}
\setcitestyle{citesep=;}

\numberwithin{equation}{section}

\usepackage{array}
\newcommand{\PreserveBackslash}[1]{\let\temp=\\#1\let\\=\temp}
\newcolumntype{C}[1]{>{\PreserveBackslash\centering}p{#1}}
\newcolumntype{R}[1]{>{\PreserveBackslash\raggedleft}p{#1}}
\newcolumntype{L}[1]{>{\PreserveBackslash\raggedright}p{#1}}

\newcommand\papertitle[1]{A new equilibrium: COVID-19 lockdowns and WFH persistence#1}
  
\title{{\Large \bf {\papertitle{}}\footnote{For helpful comments, we thank Peter Kuhn and Tim Moore. This paper uses unit record data from the Household, Income and Labour Dynamics in Australia (HILDA) Survey. The unit record data from the HILDA Survey was obtained from the Australian Data Archive, which is hosted by The Australian National University. The HILDA Survey was initiated and is funded by the Australian Government Department of Social Services (DSS) and is managed by the Melbourne Institute of Applied Economic and Social Research (Melbourne Institute). The findings and views based on the data, however, are those of the authors and should not be attributed to the Australian Government, DSS, the Melbourne Institute, the Australian Data Archive or The Australian National University and none of those entities bear any responsibility for the analysis or interpretation of the unit record data from the HILDA Survey provided by the authors.}
}\\
}
\author{
Laura Ketter\footnote{University of Queensland.}
\space\space\space\space\space\space \space\space\space\space\space\space 
Todd Morris\footnote{University of Queensland, IZA, Life Course Centre, CEPAR \& Netspar. \url{toddstuartmorris@gmail.com}}
\space\space\space\space\space\space \space\space\space\space\space\space 
Lizi Yu\footnote{University of Queensland, IZA. \url{lizi.yu@uq.edu.au}}}

\begin{document}

\maketitle \thispagestyle{empty}
\vspace{-0.3in}


\begin{abstract}
\setlength{\baselineskip}{13.5pt} 


\noindent 

\noindent This paper documents a robust link between COVID-19 lockdowns and the uptake and persistence of working from home (WFH) practices. Exploiting rich longitudinal data, we use a difference-in-differences strategy to compare office workers in three heavily locked-down Australian states to similar workers in less affected states. Locked-down workers sustain 43\% higher WFH levels through 2023~--- 0.5 days per week~--- with a monotonic dose–response relationship. Persistence is driven by adjustments on both sides of the labor market: employers downsize office space and open remote/hybrid positions, while employees relocate away from city centers and invest in home offices and technology.

\noindent 

\vspace{0.5in}


\end{abstract}

\newpage
\setcounter{page}{1}
\thispagestyle{empty}

\setcounter{tocdepth}{0}
\captionsetup[figure]{list=no}
\captionsetup[table]{list=no}

\pdfbookmark[1]{Introduction}{intro} 
\addtocontents{toc}{\protect\setcounter{tocdepth}{0}}


\noindent \textit{Nothing is so permanent as a temporary government program.}~--- Milton Friedman, 1984

\section{Introduction} \label{intro}
Emergency government interventions focus on immediate challenges but may induce lasting behavioral changes. The COVID-19 pandemic provides a rare natural experiment to examine this possibility, as governments around the world imposed sweeping lockdowns that abruptly disrupted established work patterns. While a growing body of research has documented the short-term economic effects of these policies~--- including impacts on employment, consumption, and business activity \citep[e.g.,][]{chetty2024economic}~--- much less is known about whether such emergency interventions can produce enduring changes in how people work and live.

This paper focuses on arguably the pandemic’s most enduring legacy: the widespread adoption of work from home (WFH) arrangements, which have shifted from a niche practice to a lasting feature of modern labor markets \citep{barrero2023evolution}. Despite the clear and sustained rise in remote work, the importance of public policies in driving this trend remains unclear, due to limited pre-pandemic data on WFH and difficulties in identifying appropriate counterfactuals.

\begin{figure}[htbp]
    \centering
    \caption{Lockdown timelines in Australia by state}
    \label{fig:lockdown_timeline}
    \includegraphics[width=\textwidth]{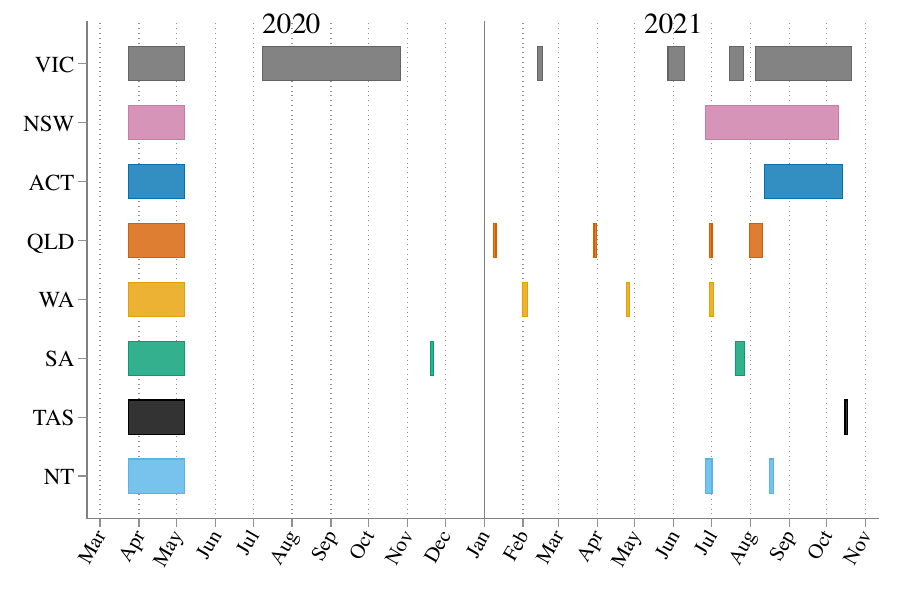}
    \includegraphics[width=0.7\textwidth]{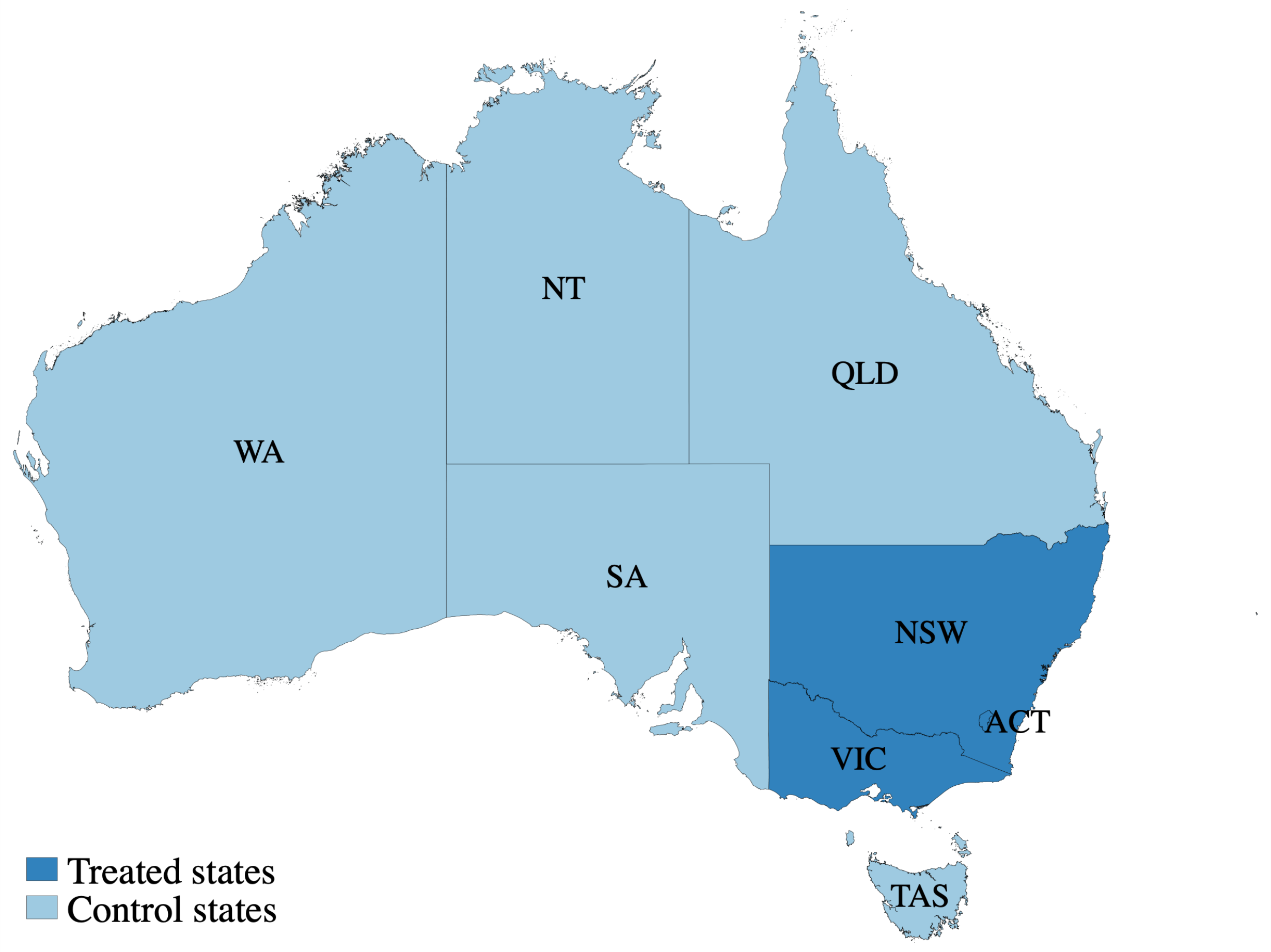}
\begin{minipage}{\textwidth}
{\footnotesize \underline{Notes}: This figure presents the timeline of lockdown periods for each state during 2020 and 2021. A state-wide lockdown is defined as any period during which stay-at-home orders were in place for the majority of the population in the state or territory, restricting movement except for essential activities. Stay-at-home orders affecting smaller portions of states' populations were rare and short-lived.} 
\end{minipage}
\end{figure}

Australia offers a compelling setting to study the impact of lockdowns on the uptake and persistence of WFH behaviors. While the country implemented a six-week national lockdown from March to May 2020, significant geographic variation in restrictions emerged thereafter. As shown in Figure \ref{fig:lockdown_timeline}, some states and territories~--- Victoria, New South Wales (NSW), and the Australian Capital Territory (ACT)~--- endured prolonged lockdowns and strict stay-at-home orders, whereas others  experienced only short and sporadic lockdowns until mass vaccination was achieved in October 2021.\footnote{Australia contains six federated states and two territories; for simplicity, we hereafter refer to all eight as ``states''. Figure \ref{fig:lockdown_timeline} provides a map of the eight states.} On average, workers in Victoria, NSW and ACT experienced 143 more days under lockdown than workers in the other five states. These geographic differences resulted mainly from localized outbreaks rather than deliberate policy decisions~--- like many Asian countries, all Australian states pursued an elimination strategy to COVID-19, under which stringent lockdowns were a standard response to local outbreaks, with lockdowns lasting until the outbreak was eliminated or mass vaccination was achieved.
 
Our study exploits the variation in lockdown duration across states~--- comparing Victoria, NSW, and the ACT to the other five states~--- in a difference-in-differences (DiD) framework, using rich individual-level data from the Household, Income and Labour Dynamics in Australia (HILDA) Survey. HILDA is a nationally representative longitudinal study that has followed thousands of Australians over time, with annual interviews conducted between August and November. It collects detailed and consistent measures of WFH since 2002 \citep{botha2023effects, lass2025working}, allowing us to construct six outcome variables capturing WFH adoption on both the intensive and extensive margins.\footnote{These variables are: (i) share of total hours worked from home; (ii) total WFH hours; (iii) indicator for any WFH; (iv) indicator for WFH under a formal agreement; (v) indicator for WFH $\geq$60\% of total hours; and (vi) indicator for fully remote work.} 
We focus on a panel of ``office workers", such as managerial, professional, or clerical roles. This group, which makes up half of the Australian workforce, was the most directly affected by stay-at-home orders during the pandemic and also the most relevant for understanding the rapid increase in WFH practices (Figure~\ref{fig:aus_wfh_share_office_nonoffice}). Our office worker sample comprises over 10,000 individuals and nearly 60,000 worker-year observations from 2002–2023. Relative to other studies, which rely on cross-sectional online surveys initiated after the pandemic began, we benefit from an existing high-quality survey with a long pre-COVID period, longitudinal structure, high response rates, and consistent WFH measures.

The long pre-COVID period and consistent measurement of WFH outcomes allow us to reliably examine the parallel-trends assumption required for the DiD estimation. In our event-study analysis, treated and control states exhibit nearly identical WFH adoption rates and trends across all six of our WFH measures, from 2002 to 2019. This parity breaks with the onset of the pandemic, as WFH outcomes rise sharply in the treated states relative to the controls, and the gap begins to narrow in 2022. 

This clear divergence motivates our formal DiD analysis to assess how lockdown duration affects WFH outcomes in the short run during the lockdown period (2020–21), and in the medium run after restrictions were lifted (2022–23).\footnote{Relatedly, prior studies exploit the extended lockdown in Melbourne, Victoria in 2020 to examine its immediate impacts on mental health \citep{butterworth2022effect}, broad human activities \citep{schurer2023quantifying}, and education outcomes \citep{gillitzer2024effect}. 
} The specification controls for state-by-occupation fixed effects, which allow WFH feasibility for the same occupation to vary across states, and survey wave fixed effects to account for national trends over time. To flexibly account for individual, job, and location characteristics, whose impacts on WFH may have changed since COVID-19 \citep{barrero2021working}, we interact these controls with indicators for the pre-COVID period (2002--19), lockdown period (2020--21), and post-COVID period (2022--23).

The estimated treatment effects are strikingly large and statistically significant at the 1\% level across all six WFH measures, indicating that the prolonged lockdowns in Victoria, NSW and ACT caused a massive and persistent uptake of remote working that would not have occurred to the same extent otherwise. On the extensive margin, workers in treated regions are 46\% (19.6 pp) more likely to engage in any WFH during the lockdown period, and 22\% (9.5 pp) more likely post-COVID. On the intensive margin, the share of hours WFH increases by 28.0 pp during the lockdown~--- a 120\% increase compared with control regions' level~--- and 10.0 pp (43\%) of this effect remains in the post-COVID period. 

To quantify the persistence of lockdown-induced WFH practices, we use the ratio of the medium-run effect (2022--23) to the immediate impact (\mbox{2020--21}). This approach provides a simple yet intuitive metric to examine how much of the initial lockdown-induced shift in WFH practices has endured. Among our outcomes, the indicator for fully remote work is the least persistent, with 11\% of the lockdown-duration effect remaining post-COVID. In contrast, the indicator for performing any WFH is the most persistent, with 49\% of the initial lockdown effect enduring post-COVID. This suggests that the expansion of WFH has largely occurred through encouraging new adopters to perform at least some work from home, rather than through sustained fully remote arrangements. This pattern is consistent with evidence that hybrid working arrangements are likely to persist in the long run, because they balance the benefits of working in the office and the benefits of WFH \citep{bloom2022hybrid, choudhury2024hybrid}. 

Our finding that lockdown duration facilitated the persistence of WFH behaviors is further supported by a positive and monotonic dose--response relationship. Across all six outcomes, the magnitudes of the WFH effect increase progressively from ACT to NSW to Victoria~--- in line with their increasing lockdown durations of 46, 90 and 206 additional days compared to control regions. For example, during the lockdown period, the WFH share increases by 11.2, 24.1 and 34.9 pp in ACT, NSW and Victoria, respectively. Post-COVID, we find no lasting effect in ACT, while workers in NSW and Victoria retain elevated WFH shares of 8.9 and 12.3 pp.\footnote{We also examine heterogeneity by occupation and worker characteristics. Consistent with previous findings \citep{bartik2020jobs, dingel2020many, adrjan2023unlocked}, we observe that occupations with higher levels of remote work pre-COVID are more likely to sustain WFH practices post-COVID, implying the importance of task feasibility and prior exposure in shaping long-term behavior.  Effects are also more pronounced among women and university-educated workers.} 

To explore potential mechanisms, we draw on complementary data sources to examine responses from both the employer and employee sides, finding evidence that adjustments by both parties have contributed to the persistent link between lockdowns and remote work. On the employer side, office vacancy rates have more than tripled in the central business districts of treated states since 2019 \citep{ramani2021donut, gupta2022work,  ramani2024working}, 
while remaining stable or even decreasing in control states, implying that firms in treated states have reduced their demand for office space. Moreover, firms in treated states  become more likely to include explicit references to WFH arrangements in new job postings, suggesting that lockdown-induced WFH practices encouraged firms to formalize WFH arrangements \citep{bamieh2022remote, hansen2023remote, adrjan2023unlocked}. On the employee side, treated-state residents display higher Google search activity for home-office infrastructure and technologies~--- such as ``computer monitor", ``office chair", and ``Zoom"~--- particularly during lockdowns. These fixed-cost investments can  lead to longer-term behavioral change by lowering barriers to continued remote work \citep{bloom2020working, davis2024work}. In addition, our longitudinal survey data reveals that office workers in treated states are more likely to move house during and after the pandemic and, conditional on moving, relocate further from the city center \citep{mondragon2022housing, akan2025new}. Taken together, these patterns suggest that both employers and employees in treated states have structurally re-optimized work arrangements to a greater extent than those in control states~--- a strategic shift that is likely to reinforce the long-term persistence of WFH \citep{barrero2021working, bick2023work, aksoy2025global}.


This paper makes two main contributions to the literature. First, we establish a robust link between lockdown intensity and WFH take-up and persistence, demonstrating that emergency policies that aim to improve health outcomes can also permanently reshape the labor market. Our differences-in-differences estimates are consistent with cross-country correlations between indices of cumulative lockdown stringency and post-COVID WFH levels \citep{aksoy2022working, 
zarate2024does}. 
However, our estimates of persistence in the Australian context are considerably larger than these associations.\footnote{We benchmark our estimates against other studies in Appendix~\ref{appendix_literature}.} In addition, our rich data and unique natural experiment allow us to document heterogeneity in the degree of persistence across different margins of WFH and by occupation, uncover several mechanisms that support persistence, and make other methodological improvements.\footnote{We benefit from a single-country setting with relatively homogeneous labor markets, similar \emph{ex ante} policies in response to COVID-19 outbreaks (i.e., immediate lockdown until the outbreak is resolved), rich longitudinal data, a causal research design, and a clearly defined measure of lockdown intensity (lockdown duration). Cross-country analyses rely on composite indices that combine different aspects of restrictiveness and enforcement \citep{brodeur2021literature}, which are harder to interpret. They also lack longitudinal data and data prior to COVID-19, and focus on a single measure of WFH.} Our results also complement studies examining regional and occupational variation in WFH adoption within countries \citep{adams2022work, bick2023work, alipour2021} by isolating the impact of lockdowns.\footnote{Our paper also relates to a growing literature examining the broader labor market implications of WFH adoption during and post the pandemic. Recent studies, which often use data from specific firms or industries, examine the impact of WFH on worker productivity \citep{gibbs2023work, emanuel2024working, lamorgese2024management}, the gender wage gap \citep{kouki2023beyond}, communication and mentoring \citep{emanuel2023power}, innovation \citep{brucks2022virtual}, entrepreneurship \citep{kwan2025entrepreneurial}, and other labor market outcomes. Our study provides an empirical framework that can be extended to consider the implications of WFH among a large and representative set of workers and firms.}


Second, this paper helps bridge the growing body of research on the short-term disruptions of the pandemic with emerging concerns on its longer-term consequences \citep{jorda2022longer}. Major life events or shocks can durably shape individual preferences and behaviors \citep{malmendier2011depression, callen2015catastrophes, severen2022formative}. In the case of COVID-19, while a large literature has documented immediate behavioral shifts during the pandemic \citep[e.g.,][]{barrero2021working, brynjolfsson2020covid}, less is known about the extent to which these changes persisted once restrictions were lifted and the longer-term effects of lockdowns \citep{berkes2025lockdowns}. We provide rare longitudinal evidence that~--- in the context of WFH~---  the impacts of lockdowns endure due to a range of responses from employers and employees. Changes in technology adoption, office usage, hiring practices, and commuting distances are all examples of lockdown-induced structural shifts that may facilitate path dependence \citep{davis2024work, kaplan2020great}. Our findings also relate to the broader economics literature on habit formation, which considers the longer-term impacts of temporary interventions on a range of important behaviors,\footnote{Studies include interventions affecting smoking cessation \citep{gine2010put}, consumption of utilities \citep{costa2021hysteresis, byrne2024nudges, ito2018moral, ferraro2013using}, physical exercise \citep{charness2009incentives, royer2015incentives}, commuting behavior \citep{larcom2017benefits}, political participation \citep{fujiwara2016habit}, road safety \citep{moore2024shaping}, and hospital hygiene \citep{steinywellsjo2025}.} but not where and how people work.
In our context, the COVID-19 pandemic can be seen as a generation-defining event that introduced new routines, expectations, and investments. As these structural changes reshaped work habits, a new equilibrium emerged.



The rest of the paper is organized as follows. Section \ref{section2} describes the policy background and our data sources. Section \ref{section3} outlines the empirical strategy, and Section \ref{section4} presents the main results. Section \ref{section5} explores potential mechanisms. Section \ref{section6} presents additional robustness checks and describes heterogeneity in the effects. Section \ref{section8} concludes.

\addtocontents{toc}{\protect\setcounter{tocdepth}{0}}
\pdfbookmark[1]{Background and Data}{section2} 
\section{Background and Data} \label{section2}
In this section, we first provide background on regional variations in COVID-19 evolution and lockdown policies across Australia, and then describe the dataset we use in our empirical analysis and the construction of our six WFH outcome measures.

\pdfbookmark[2]{COVID-19 in Australia}{COVID-19background} 
\subsection{COVID-19 in Australia}
\begin{figure}[htbp]
\centering
\caption{Evolution of COVID-19 pandemic in Australia} \label{fig:covid_aus}
\begin{subfigure}[t]{0.49\textwidth}
\caption{Confirmed new cases: Aus/US}
\includegraphics[width=\textwidth]{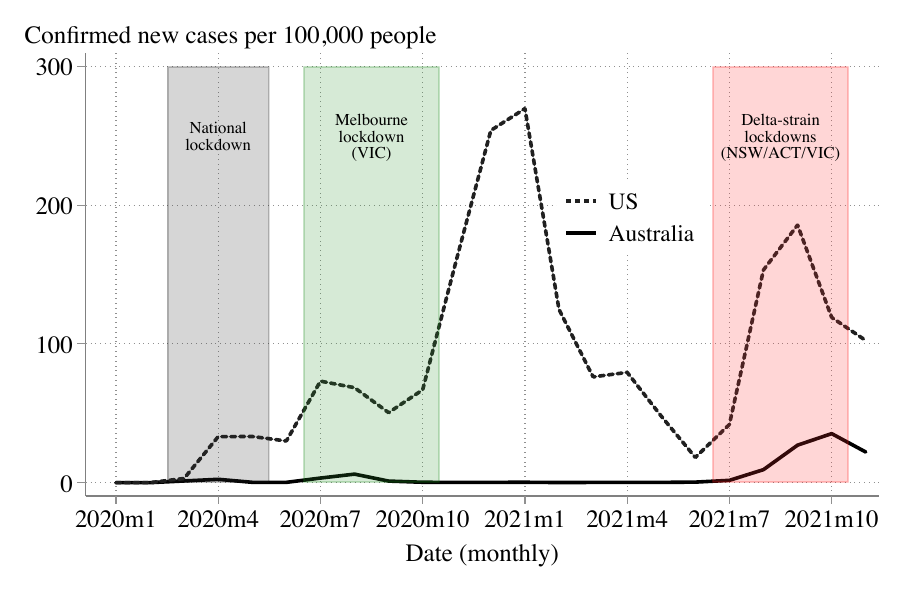} 
\end{subfigure}
\begin{subfigure}[t]{0.49\textwidth}
\caption{Cumulative excess deaths: Aus/US/World}
\includegraphics[width=\textwidth]{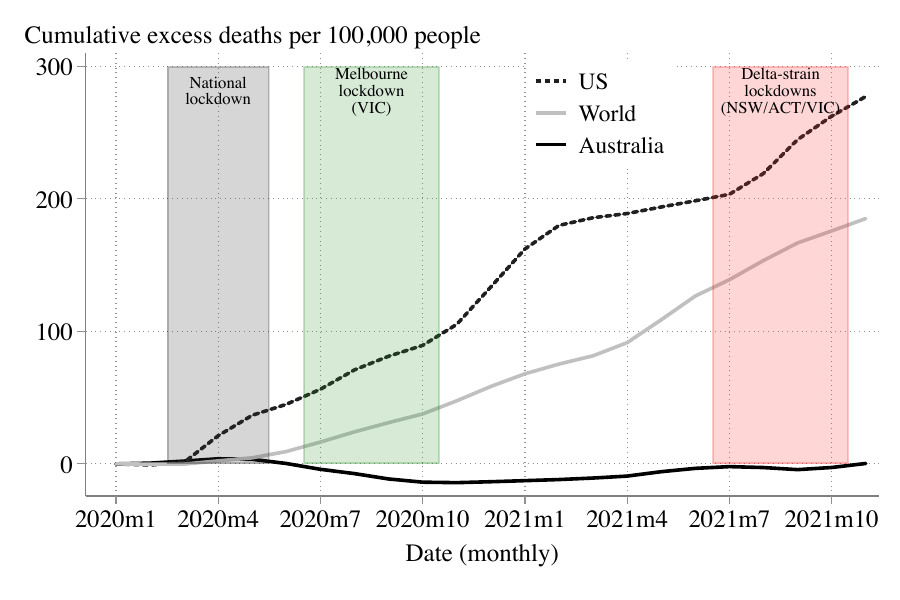}
\end{subfigure} \\ \vspace{1cm}

\begin{subfigure}[t]{0.49\textwidth}
\caption{Confirmed new cases by state}
\includegraphics[width=\textwidth]{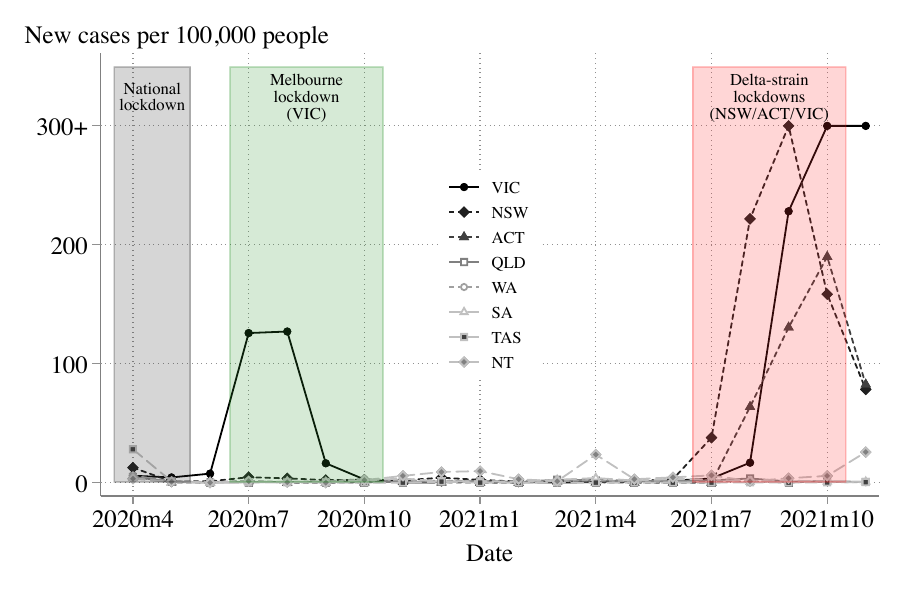} 
\end{subfigure}
\begin{subfigure}[t]{0.49\textwidth}
\caption{Change in workplace mobility by state}
\includegraphics[width=\textwidth]{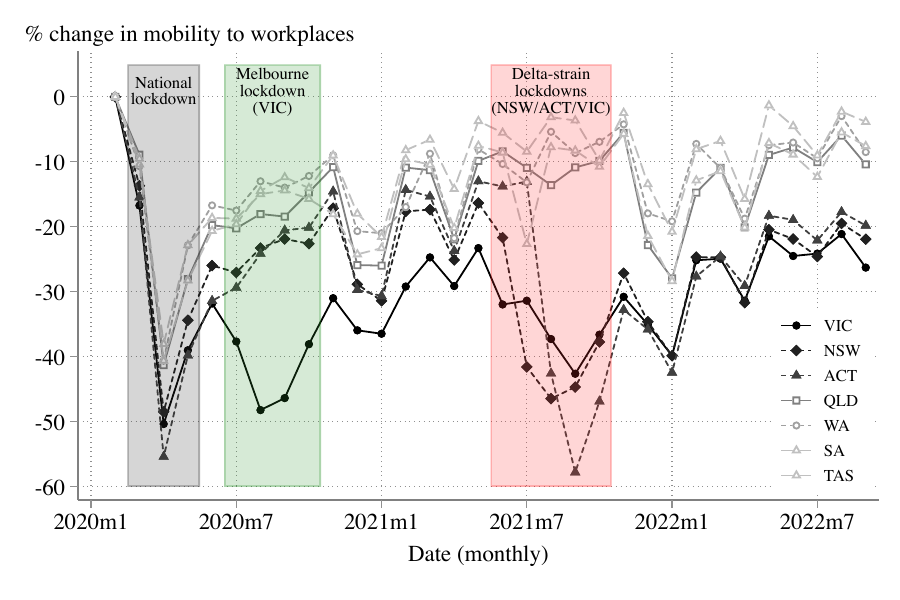}
\end{subfigure}



\begin{minipage}{\textwidth}
{\footnotesize \underline{Notes}: These figures plot the evolution of the COVID-19 pandemic in Australia. The top two panels (a) and (b) compare the trajectory of the pandemic in Australia and the United States, using measures of confirmed case counts and excess deaths. The bottom two panels (c) and (d) show the variation in pandemic experiences across Australia’s eight states and territories. The three shaded periods correspond to the National lockdown, the Melbourne (Victoria) lockdown and the Delta-strain lockdown, which started in NSW and led to lockdowns in Victoria and the ACT.} 
\end{minipage}
\end{figure}

The COVID-19 pandemic had a profound global impact, and Australia's experience differed significantly from that of the United States and most Western countries. From 2020 to 2021, Australia managed the pandemic with strict border controls, lockdowns, and targeted public health measures that collectively aimed for COVID-zero. This approach resulted in markedly fewer confirmed cases compared to the United States. Figure \ref{fig:covid_aus}(a) illustrates that throughout the pandemic, Australia’s confirmed new cases per 100,000 people remained significantly lower than those in the US. By December 2021, Australia had recorded just 1,399 cumulative cases per 100,000 people, compared to 15,675 cases per 100,000 in the US, despite similar rates of testing per capita and a much smaller share of positive tests.\footnote{Source: Our World in Data, \textit{Coronavirus (COVID-19) Cases.} \url{https://ourworldindata.org/covid-cases}} Consequently, Australia saw no excess deaths during this period, as shown in Figure \ref{fig:covid_aus}(b), whereas the United States experienced a cumulative total of nearly 300 excess deaths per 100,000 people.

This stark contrast in outcomes is largely attributed to Australia's swift and stringent measures (as well as natural advantages, such as being a relatively remote island nation). On March 13 2020, a national cabinet was convened, comprising all of the key state and federal decision-makers: Australia's prime minister and all eight state premiers.\footnote{Leaders of the two territories are called `chief ministers' rather than premiers. For simplicity, we refer to the territories as `states' throughout the paper, as territories are autonomous regions that made decisions regarding lockdowns and other COVID-19 mitigation policies.} This reflected a desire for the country to have a unified approach to the threat of the COVID-19 virus. As case numbers started to rise towards the end of March, all state governments implemented similar lockdowns (indicated by the gray-shaded area in Figure \ref{fig:covid_aus}), asking people to work and study from home, closing non-essential businesses, banning unnecessary domestic and international travel, and prohibiting mass gatherings. Australian borders were also closed to non-residents and two weeks of hotel quarantine was instituted for all international arrivals. High levels of testing, contact-tracing and quarantine procedures for potential cases also helped prevent the transmission of the virus. As case numbers declined to very low levels by May 2020, restrictions were eased. Thereafter, the eight Australian states/territories followed extremely different COVID-19 trajectories (Figure \ref{fig:covid_aus}(c)) and lockdown policies (Figure \ref{fig:lockdown_timeline}) due to local outbreaks.


In July 2020, a single breach in Melbourne’s hotel quarantine program led to an outbreak across the state of Victoria, forcing Greater Melbourne to endure one of the world's longest lockdowns that year~--- 112 days (represented by the green-shaded area).\footnote{Regional Victoria also experienced lockdowns during this period, but restrictions were shorter and less stringent, as the virus spread mainly within Greater Melbourne (where three-quarters of Victorians live).} This lockdown was not only much longer than the initial national lockdown but also more stringent.\footnote{For example, a ban on being more than 5 km from home was introduced, playgrounds were closed, outdoor exercise was limited to one hour per day, and a nighttime curfew (8pm--5am) was instituted.} The long and stringent lockdown eventually achieved its goal, suppressing the virus back to zero in Victoria, and other states managed to remain at zero by closing their borders to Victorians. Infection levels remained minimal nationwide until mid-2021, when the highly transmissible Delta variant emerged as the dominant strain in many countries. During the Delta outbreak, NSW became the epicenter, prompting a long and stringent lockdown in Greater Sydney.\footnote{Other parts of NSW were also subject to lockdowns, but to a lesser extent. Around two-thirds of NSW residents live in Greater Sydney.} Victoria and the ACT also experienced Delta waves, resulting in more extended lockdowns (denoted by the red-shaded area). In contrast, other states managed to maintain low case numbers throughout the same period without lockdowns.
 
Google mobility data presented in Figure \ref{fig:covid_aus}(d) illustrate how Australians adapted to lockdowns and restrictions, as well as the lasting effects of lockdowns on mobility patterns. Compared to pre-pandemic levels, workplace mobility declined significantly by 40--50\% across all states during the initial national restrictions, with some modest differences across states, likely due to differences in the share of office workers. Victoria, which faced the strictest and longest lockdown in 2020, saw workplace mobility drop by 35--50\%, while other states gradually rebounded as restrictions eased. By June 2021, some states had nearly returned to pre-pandemic mobility levels, whereas Victoria’s workplace mobility remained 20\% lower than before the pandemic. 

During mid-2021, the Delta outbreak and the following lockdowns in NSW, Victoria, and ACT triggered another sharp decline in mobility in these three states, with reductions of 30--60\%. In contrast, other states experienced minimal mobility declines during this period. Even after restrictions were lifted, workplace mobility in the three locked-down states (NSW, Victoria, and ACT) remained lower than in regions that avoided prolonged lockdowns up until the Google Mobility data was discontinued (in mid-October 2022).

\pdfbookmark[2]{Data and WFH Measures}{data} 
\subsection{Data and WFH Measures}
Our main analysis is based on an unbalanced panel of respondents to the Household Income and Labour Dynamics in Australia (HILDA) Survey. HILDA is a large, nationally representative, household‐based longitudinal survey that started in 2001 and has since spawned a large body of research \citep{watson2012hilda}. Each year, over 20,000 Australians are interviewed, with the overwhelming majority of interviews completed between August and November each year. Individuals are re-interviewed indefinitely with much less attrition than other household-level longitudinal surveys \citep{watson2018best}.\footnote{Prior to COVID-19, around 93\% of interviews were face-to-face, with telephone interviews (7\%) used as a last resort. In 2020 and 2021, telephone was the primary interview mode due to social distancing requirements (96\% and 76\% in the respective years), but the timing and content of the survey remained similar, except for the addition of a special `Coronavirus' module in 2020. The share of telephone interviews has since declined but remains higher than pre-pandemic (28\% in 2022 and 21\% in 2023).} In each wave, detailed information is collected on each member of the household, including information on their income, employment, hours worked, job characteristics, education, relationships, and health. 


Importantly, HILDA has consistently collected information on WFH practices since wave 2 (2002).\footnote{We measure WFH based on three main questions: (i) total work hours in main job: ``including any paid or unpaid overtime, how many hours per week do you usually work in your main job?'', (ii) any WFH: ``in your main job, are any of your usual working hours worked at your home?''; (iii) WFH hours: ``approximately how many hours each week do you usually work at home for your main job only?''. Question (iii) is a follow-up question for those who respond `yes' to question (ii). } We derive six key outcome variables indicating different aspects of remote work for a worker's main job: (i) the \textbf{WFH share} ~--- the fraction of a worker's total weekly hours that are usually done from home; (ii) the total number of \textbf{WFH hours} per week; (iii) an indicator for \textbf{any WFH}; (iv) an indicator for \textbf{WFH under a formal agreement} with the employer; (v) an indicator for \textbf{mostly WFH} (working at least 60\% of hours from home, roughly equivalent to 3+ days per week); and (vi) an indicator for \textbf{fully remote} work (i.e., WFH share of 100\%).\footnote{We winsorize the total number of hours worked and the number of WFH hours at 75, which is the 99th percentile for total hours worked. Just 0.1\% of workers report working more than 75 hours from home~--- these individuals are classified as fully remote.}

We use survey waves 2002–2023, covering the pre-pandemic and pandemic periods, and we restrict the sample to working-age adults (18–59), who are employees (excluding the self-employed), living in major urban areas (cities with at least 100,000 residents), and employed in `office jobs'.\footnote{Approximately 51\% of workers in the sample are employed in office jobs, primarily comprising clerical and administrative workers, managers, and professionals. The office job category also includes selected occupations from other groups, such as Engineering, ICT and Science Technicians; Other Technicians and Trades Workers; and Sales Representatives and Agents. Non-office jobs are mainly made up of community and personal service workers (23\%), technicians and trades workers (16\%), sales workers (13\%), laborers (13\%), education professionals (13\%), health professionals (10\%), and machinery operators and drivers (10\%).} Focusing on major urban cities makes the treated and control regions more comparable in terms of job mix, population density, and labor market norms, and excludes rural areas that saw very few COVID cases or lockdowns. We focus on workers employed in office jobs because this group was the most affected by remote work mandates during the pandemic and this focus further improves the comparability of workers in treated and control regions.\footnote{Figure~\ref{fig:aus_wfh_share_office_nonoffice} compares the evolution of WFH in Australia for office and non-office workers. Among office workers, WFH only rise gradually from 5\% in 2002 to 8\% in 2019, consistent with evidence from the US \citep{harrington2023has}, surged past 40\% during the pandemic, and settled at 31\% by 2023. For non-office workers, WFH also rose during COVID-19 but peaked at just 9\% in 2021 and quickly declined thereafter.} 


As presented in Table~\ref{tab: summarystats}, our main analysis sample includes 57,896 individual-year observations from 10,127 unique individuals. On average, workers in the sample are 38 years old and approximately half are female. Two-thirds are partnered (71\%), 46\% have a university degree, and they work 39 hours per week on average. Moreover, around 32\% of office workers engaged in some form of WFH, with WFH hours accounting for 12\% of their total working hours. About 63\% of the sample are from workers residing in treated regions (Victoria, NSW, and ACT).

\pdfbookmark[2]{Supplementary Data Sources}{Supplementary} 
\subsection{Supplementary Data Sources}
In addition to the individual-level microdata from the HILDA Survey, we draw on several supplementary data sources to examine employer and employee responses to lockdown-induced shifts in work arrangements. These additional datasets allow us to explore potential mechanisms contributing to the persistence of WFH and to validate patterns observed in our main analysis.

\noindent \textbf{Office Vacancy Rates.} To capture employer-side adjustments in the demand for physical office space, we use CBD office vacancy rate data compiled by the Property Council of Australia. These data are reported twice a year and cover vacancy rates across major central business districts in all of Australia's state capitals. Unlike in the US, where state capitals are relatively small cities, all of Australia's state capitals are by far the state's largest city and comprise a large share of the state's population.

\noindent \textbf{New Job Posting Data.} To assess changes in how employers advertise remote work opportunities, we draw on WFH-specific job posting data from \href{https://wfhmap.com/}{WFHmap.com}, as used by \cite{hansen2023remote}. This database uses natural language processing to extract WFH-related language from the near-universe of online job ads across English-speaking countries, including Australia. Each job ad is classified by an algorithm developed by \cite{hansen2023remote} as either `explicitly allowing WFH' or `not explicitly allowing WFH'. Our custom data~--- graciously provided by the authors~--- contains, for 2019--24, the total number of job ads and the total number of WFH job ads in Australia for each combination of the following variables: state, subregion (SA4 statistical area), calendar month and year, two-digit occupation and two-digit industry. We restrict our focus to ads for `office jobs' in major urban areas, to match our HILDA sample, using the information on subregion and occupation.

\noindent \textbf{Google Search Activity.} To capture employee-side investment and adaptation, we collect monthly Google Trends data on search interest in home-office-related keywords such as ``computer monitor", ``office chair”, and ``Zoom”. These indicators reflect consumer interest in remote work infrastructure and are available monthly at the state level.

\pdfbookmark[1]{Empirical Strategy}{section3} 
\section{Empirical Strategy} \label{section3}
We examine how the differential exposure to COVID-19 lockdowns across Australian states affected the uptake of WFH by office employees in both the short run and the medium run. Our empirical approach is a difference-in-differences design that exploits the quasi-experimental variation in lockdown stringency between treated and control regions.\footnote{Similar research designs have been used to look at the impact of lockdowns on other outcomes during the pandemic, including mental health \citep{butterworth2022effect, schurer2023quantifying} and children's test scores \citep{gillitzer2024effect}.} 
Specifically, we define the treated group as office workers in major urban areas in the states that experienced extensive lockdowns in 2020–21 (Victoria, NSW, and the ACT) and the control group as office workers in the other states and territories. We compare WFH outcomes between these two groups before, during, and after the pandemic to isolate the causal impact of the lockdown-induced WFH shock. Specifically, we estimate regressions of the following form: 
\begin{equation}
\label{eq:basicDiD}
y_{iost} = \beta_{0} +  \beta_{1}(\text{Treat}_{ist} \times \text{Lockdown}_{t}) +  \beta_{2}(\text{Treat}_{ist} \times \text{Post}_{t})+ X'_{iost} \mathbf{\gamma} + \lambda_t + \delta_{os} +\varepsilon_{iost}, 
\end{equation}
where $y_{iost}$ is a WFH outcome for individual $i$ of occupation $o$ who resides in state $s$ in year $t$. $\text{Treat}_{ist}$ is an indicator for individual $i$ residing in one of the treated regions (Victoria/NSW/ACT), $\text{Lockdown}_t$ is an indicator for the years 2020--21, and $\text{Post}_t$ is an indicator for 2022--23.\footnote{We use an individual's current state of residence to assign treatment. Interstate migration in our sample was low during this period, with only 2.7\% of individuals moving from treated to control states, or vice versa, from 2019 to 2022. Our robustness check using 2019 location to assign treatment is presented in Figure \ref{fig: robust} and we find nearly identical estimates.} The coefficients $\beta_1$ and $\beta_2$ are of primary interest: $\beta_1$ captures the immediate impact of the average treatment effect on WFH during the lockdown period, and $\beta_2$ captures the medium-run effect in the post-pandemic period. Moreover, the ratio $\beta_2/\beta_1$ can serve as a summary measure of persistence, with higher values indicating that a larger share of the initial WFH increase during lockdown persisted into the post-COVID era. 

We include a rich set of controls $X_{iost}$ to account for other factors influencing WFH adoption. These controls include individual demographic characteristics (age, gender, marital status, educational attainment, and the presence of dependent children), job characteristics (years of job tenure, a full set of industry dummies at the 4-digit level, and occupation dummies at the 2-digit level), and a location proxy for commuting costs (the average commuting time in 2019 for full-time workers living in the same postcode as individual $i$). To allow these factors to have different relationships with WFH over time, each control is interacted with period indicators for the pre-pandemic (2002--2019), lockdown (2020--21), and post-pandemic (2022--23) periods. This flexible specification permits, for example, the effect of having young children on WFH to vary before and after COVID. 

We also include survey-year fixed effects $\lambda_t$ to capture common time shocks, such as nationwide technological trends and changes in social norms. To absorb static differences in WFH propensity across regions and occupations, we include state-by-occupation fixed effects $\delta_{os}$. These fixed effects, defined for each combination of state $s$ and one-digit occupation category $o$, control for any time-invariant factors that are specific to a state and occupation pair. For instance, NSW-based professionals might generally have higher WFH rates than QLD-based professionals, regardless of the pandemic. By controlling for $\delta_{os}$, we ensure that identification comes from deviations from those baseline state-occupation differences. Correspondingly, standard errors are clustered at the state-by-occupation level.\footnote{As shown in Figure \ref{fig: robust}, our results are robust to alternative clustering choices (e.g., clustering by state or by individual) and to using wild cluster-robust bootstrap procedures to accommodate the limited number of state-level clusters \citep{cameron2008bootstrap}. We also obtain robust results when using alternative fixed effects structures and when defining the treatment/control groups based on individuals' state of residence in 2019 or using alternative samples not restricted to office workers only.}

Our DiD method relies on two key assumptions: (i) workers in the treatment group were induced to WFH at higher rates due to extended local lockdown restrictions and (ii) parallel-trends~--- WFH behaviors would have evolved in the same way in the treatment and control groups if not for the differential impacts of COVID-19 across states. Figure~\ref{fig: raw_wfh_share_treat_control}  provides strong empirical support for both assumptions. Before the pandemic, the WFH share was extremely similar in the treatment and control groups, both in its level and trend. Then, in 2020 and 2021, we observe a much larger increase in the treated states~--- the WFH share increased sevenfold from 2019 to 2021 in the treated states (8\% to 59\%), with a much smaller increase in other states (9\% to 23\%). The graph also suggests that these large differences across states persist through to 2023, albeit with some attenuation from 2021 to 2022. We formally confirm this by estimating an event-study model that allows for dynamic treatment effects in each year relative to 2019. These estimates (discussed in detail in Section~4.2) show no significant differences between treated and control states in any pre-pandemic year, across all our WFH measures. 

\pdfbookmark[1]{Main Results}{section4} 
\section{Main Results} \label{section4}
\pdfbookmark[2]{Baseline Difference-in-Differences Analysis}{DiD} 
\subsection{Baseline Difference-in-Differences Analysis}
Table \ref{tab: did} summarizes the DiD estimates for all six WFH outcome variables in the lockdown period and the post-pandemic period. Notably, the two specifications in Panels A and B yield very similar estimates. Including detailed period-specific controls $X_{iost}$ in Panel B only marginally attenuates the coefficients, leaving the major pattern unchanged. This suggests that the observed WFH increase in treated areas is not driven by other coincident factors or pre-existing variations across states, but rather by the much larger uptake of WFH in the treated states due to lockdowns in response to localized outbreaks.

\begin{figure}[htbp]
\centering
\caption{Trends in WFH share for office workers in treatment/control regions} \label{fig: raw_wfh_share_treat_control}
\includegraphics[width=\textwidth]{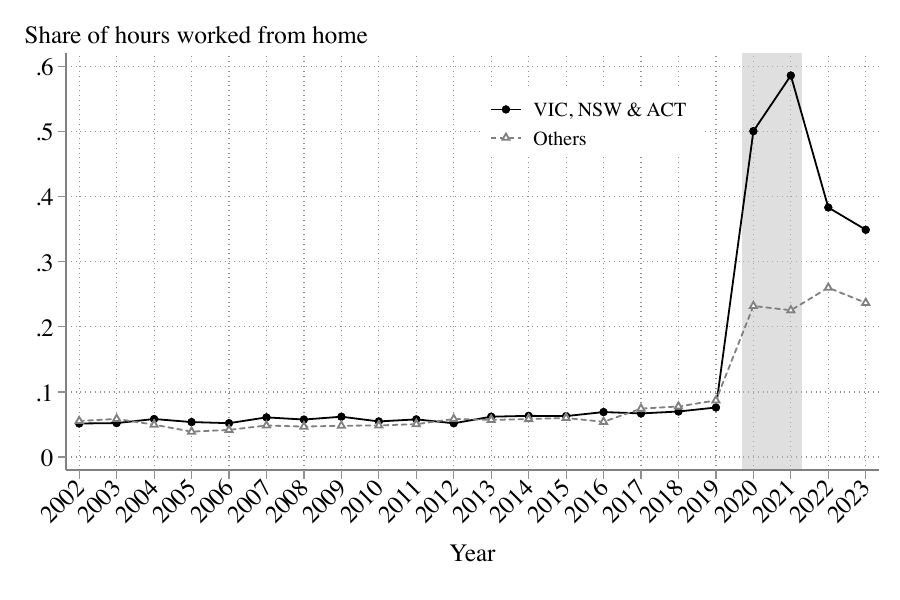}

\begin{minipage}{\textwidth}
{\footnotesize \underline{Notes}: This figure plots the share of WFH during the analysis period for office workers in the major urban cities, for the treated states (VIC, NSW and ACT) and control states separately. WFH share is measured as the proportion of hours worked from home relative to total hours worked in the main job. Office workers are primarily composed of clerical and administrative workers, managers, and professionals. } 
\end{minipage}
\end{figure}

Focusing first on the lockdown period (2020--21), we observe a dramatic uptake of WFH among office workers in the treated states relative to those in control states. All estimated $\beta_1$ coefficients are large in magnitude and significant at the 1\% level. For instance, the WFH share~--- the fraction of total hours worked from home~--- was about 28.0 pp higher in the treated states than in the controls. To assess the magnitude of these effects, we use the control states' average level in 2020 as the baseline.  Because control states also experienced the national lockdown, their 2020 level reflects the initial increase in WFH due to the pandemic, making it a reasonable benchmark to compare the additional impact of extended lockdowns in the treated states.\footnote{We also observe minimal change in the control states average WFH levels from 2020 to 2023 (Figure~\ref{fig: raw_wfh_share_treat_control}), which further verifies the usefulness of their 2020 level as a benchmark.} Relative to the baseline level of 23.4\%, the estimated coefficient of 28.0 pp represents a substantial increase of 120 percent. Consistent with this finding, treated-state workers work 11.0 more hours per week from home than control-state workers during the lockdown period, representing a 126 percent increase. On the extensive margin, the likelihood of working any hours from home is 19.6 pp (46\%) higher in treated states than in control states. We also find that workers in treated states are 21.3 pp (68\%) more likely to WFH with a formal agreement with their employer during the lockdown, 30.8 pp (164\%) more likely to be mostly WFH, and 29.1 pp (227\%) more likely to be fully remote. These effects confirm that the strict lockdowns have induced a massive shift toward remote work in the treated regions. 

Turning to the post-pandemic period 2022--23, we find that a significant portion of the WFH increase in treated states persists even after all restrictions were lifted. The DiD estimates $\beta_2$ remain positive and statistically significant (at the 1\% level) for all six outcomes. As expected, the magnitudes are smaller than during the lockdown years, but they are still economically large. On the extensive margin, the treated-control gap in the probability of any WFH is about 9.5 pp in the post-pandemic period~--- roughly 49\% of the corresponding increase during 2020--21. Relatedly, 48\% of the rise in WFH or hybrid arrangements during the lockdown period persists into post-COVID period. In contrast, fully remote work shows much less persistence: the treated-control difference in the fully-WFH indicator drops to roughly 3.1 pp post-pandemic, only about 11\% of its lockdown-period level. This indicates that most workers who had been entirely remote eventually returned to the office at least some of the time. 

Other facets of WFH measures also exhibit considerable persistence, on the intensive margin. The treated-state premium in the WFH share is around 10.0 pp in 2022–23, which is about 36\% of the lockdown-period effect. Workers in treated states continue to log  more hours from home, and continue to work mainly from home (majority of the week) in 2022–23. On average across these metrics, we estimate that the medium-run WFH levels in treated states are around one-third of the initial increases observed during the lockdown period. This persistence ratio underscores that the COVID lockdowns have led to not just a temporary spike, but a lasting shift in work behavior even after the emergency conditions waned. 

Notably, the persistence is concentrated in hybrid work rather than full-time remote work. While most employees who became fully remote eventually return to the office at least part-time, a large number continue to WFH one or more days per week. This is consistent with anecdotal and survey evidence that a hybrid model becomes the preferred long-term equilibrium for many workplaces, balancing the benefits of face-to-face collaboration with the flexibility of WFH \citep{bloom2022hybrid}.

\pdfbookmark[2]{Dynamic DiD Analysis and WFH Trajectories}{eventstudy} 
\subsection{Dynamic DiD Analysis and WFH Trajectories}
Next, we formally test the parallel trends assumption using a dynamic DiD analysis by estimating the following specification:
\begin{equation}
\label{eq:eventstudy}
y_{iost} = \beta_{0} + \sum_{\substack{k=2002 \\ k\neq 2019}}^{2023} \beta_{k} \text{Treat}_{isk} \times \mathbf{1}(t=k) + X'_{iost} \mathbf{\gamma} + \lambda_t + \delta_{os} +\varepsilon_{iost},
\end{equation}
where the key coefficients of interest are the $\beta_k$ terms. These coefficients capture the difference in WFH behaviors in year $k$ between workers in treated and control states, relative to their difference in the pre-pandemic year of 2019. All other variables are defined as in Equation \ref{eq:basicDiD}.

\begin{figure}[htbp]
\centering
\caption{Dynamic difference-in-differences estimates relative to 2019} \label{fig: eventstudy}

\begin{subfigure}[t]{0.49\textwidth}
\caption{WFH share}
\includegraphics[width=\textwidth]{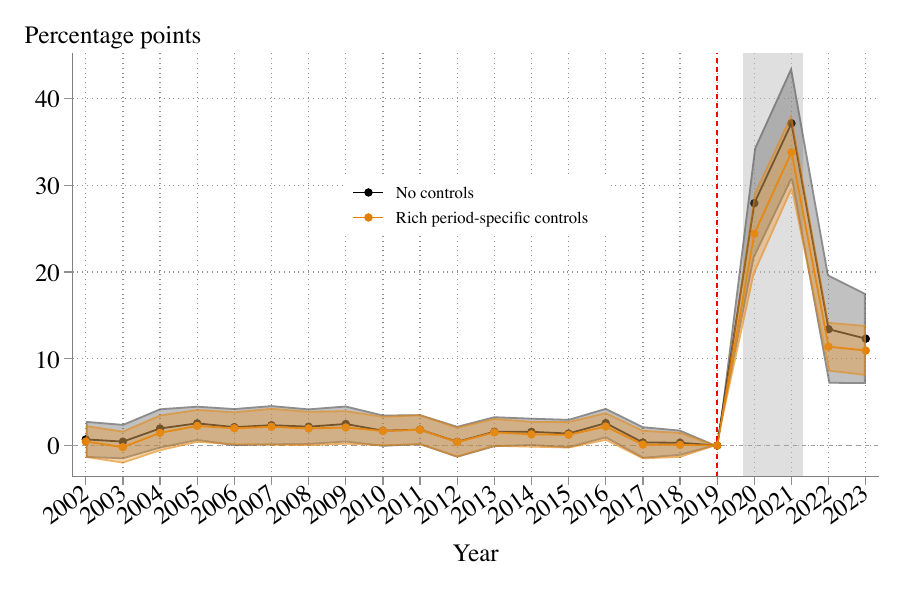}
\end{subfigure}
\begin{subfigure}[t]{0.49\textwidth}
\caption{WFH hours per week}
\includegraphics[width=\textwidth]{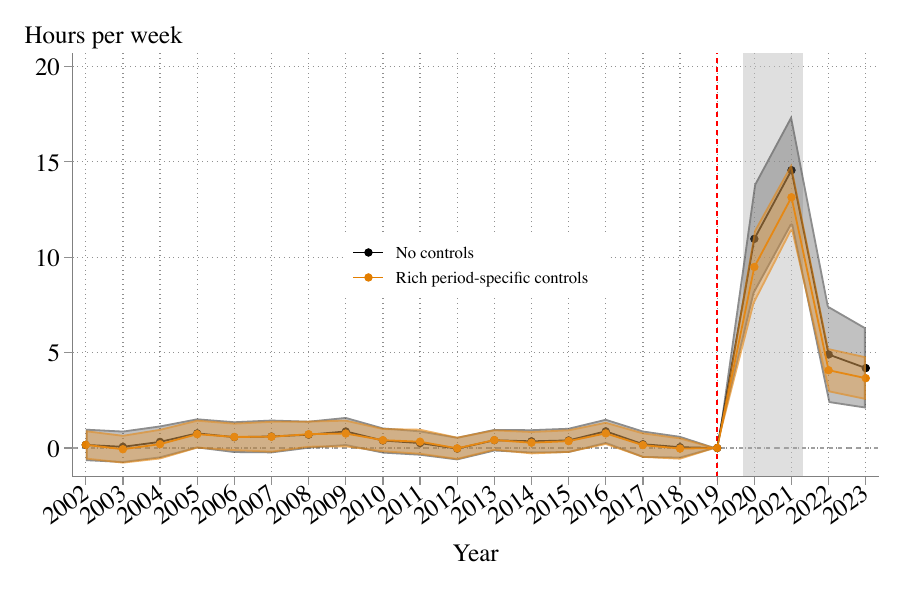} 
\end{subfigure} \vspace{0.1cm}

\begin{subfigure}[t]{0.49\textwidth}
\caption{Any WFH}
\includegraphics[width=\textwidth]{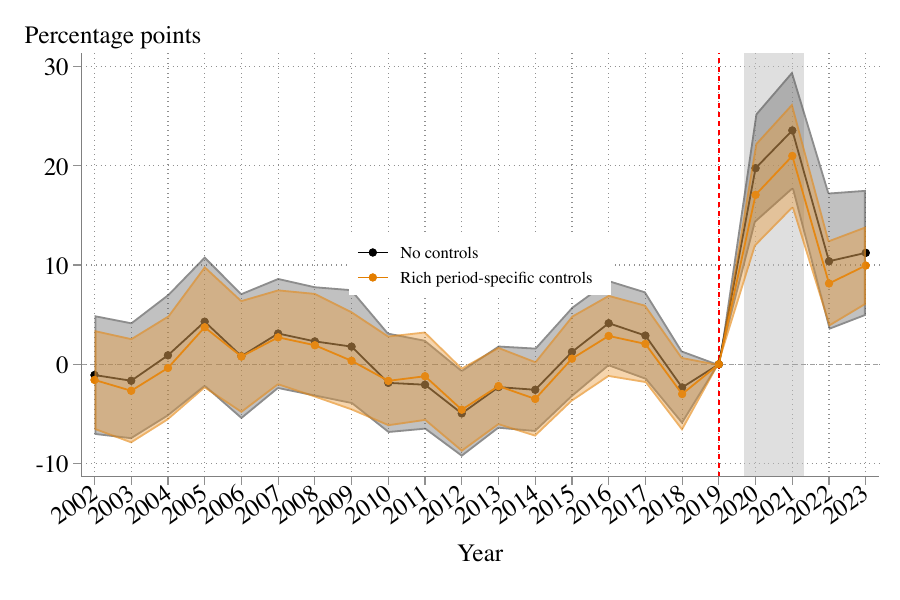} 
\end{subfigure}
\begin{subfigure}[t]{0.49\textwidth}
\caption{Any WFH with formal agreement}
\includegraphics[width=\textwidth]{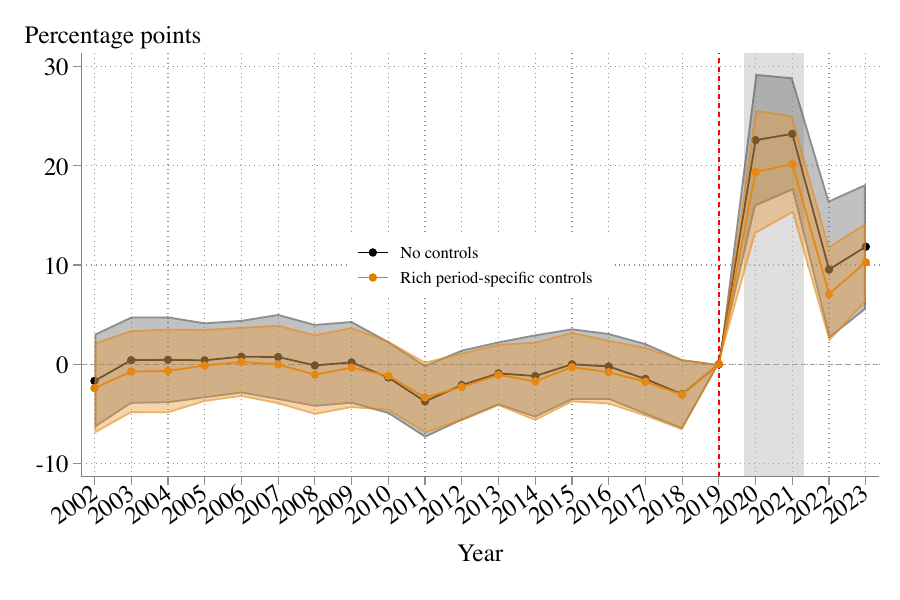} 
\end{subfigure} \vspace{0.1cm}

\begin{subfigure}[t]{0.49\textwidth}
\caption{Mostly WFH (WFH share $\geq$ 0.6)}
\includegraphics[width=\textwidth]{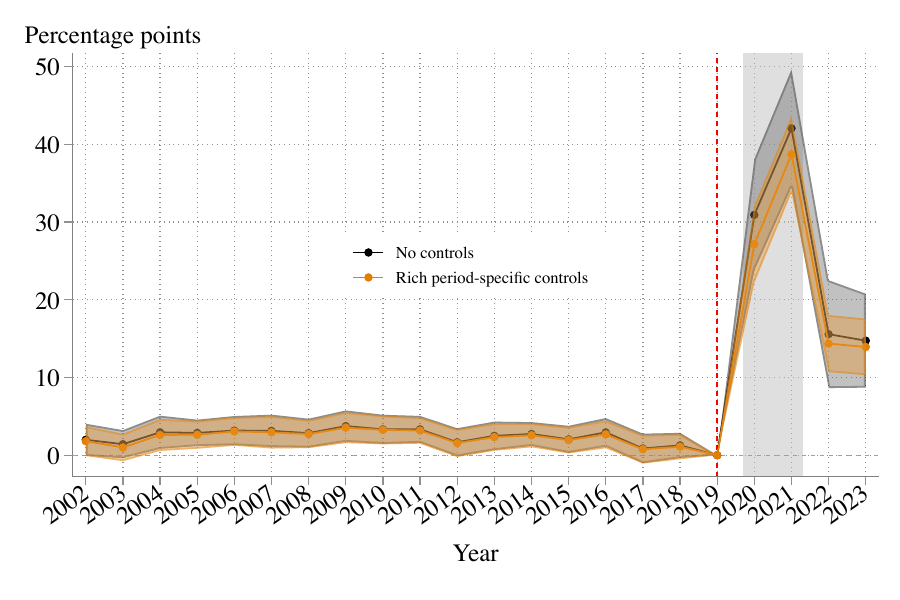}
\end{subfigure}
\begin{subfigure}[t]{0.49\textwidth}
\caption{Fully remote (WFH share $=$ 1)}
\includegraphics[width=\textwidth]{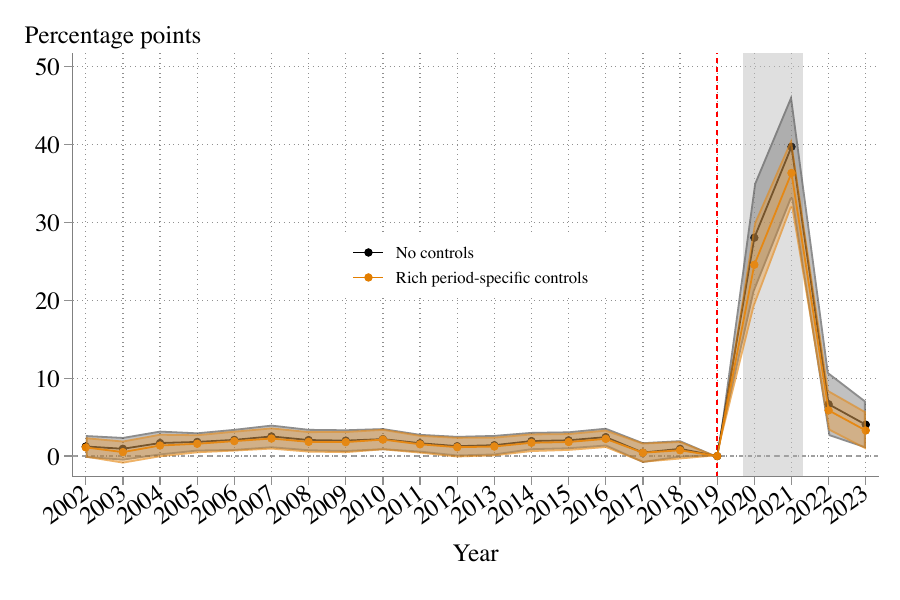} 
\end{subfigure}

\begin{minipage}{\textwidth}
{\footnotesize \underline{Notes}: These figures present the event-study estimates, along with 95\% confidence intervals, for the six WFH measures, based on Specification \eqref{eq:eventstudy}. The gray line shows estimates controlling for fixed effects only, while the orange line shows estimates additionally controlling for period-specific individual, job, and location characteristics.} 
\end{minipage}
\end{figure}

Figure \ref{fig: eventstudy} presents the estimated coefficients $\beta_k$ from estimating the previous specification on the six WFH outcomes, along with their 95\% confidence intervals. The two specifications (with and without rich period-specific controls) yield very similar trajectories, suggesting that the observed WFH increase in treated areas is mostly driven by the much larger uptake of WFH in the treated states.

Before 2019, the estimates are consistently close to zero and display little evidence of any trend, which confirms our parallel-trends assumption. At the onset of the pandemic in 2020, a sharp divergence emerges. The treated states ~--- which implemented stringent lockdowns ~--- experience an immediate and pronounced increase in remote work relative to the control states. These differences continue into 2021, with point estimates reaching their peak around that year. In 2022 and 2023, the estimates are considerably smaller, but still positive and economically large. Interestingly, there is little difference between the 2022 and 2023 estimates, particularly for hybrid WFH arrangements, indicating that the higher rates of WFH in the treated states are likely to endure,\footnote{While HILDA data beyond 2023 is not yet available, our comprehensive job postings data show a sustained positive effect on the share of job ads that explicitly allow WFH in treated states relative to control states through to the end of 2024, with no sign of decay.} while the effects on fully-remote work appear set to decline further. 

\pdfbookmark[2]{Dose-Response Relationship}{dose} 
\subsection{Dose-Response Relationship}
Even among the treated states~--- Victoria, NSW, and the ACT~--- there still exists considerable variation in the intensity and duration of each state's lockdown policies. In particular, during the lockdown period in 2020--21, the ACT experience only 46 more days of lockdown than the average among the control states, while NSW and Victoria experience 90 and 206 more lockdown days. To formally assess how WFH outcomes vary with lockdown exposure, we modify our baseline DiD model (Specification~\ref{eq:basicDiD}) by allowing for different short- and medium-term treatment effects in each of the three treated states.

We present the estimates graphically in Figure~\ref{fig: dose}. On the $y$-axis, we present our point estimates and 95\% confidence intervals, while the points on the $x$-axis for each state reflect the intensity of the treatment: the additional days of lockdown relative to the control group. Across all six outcome measures, we observe several strikingly consistent and robust patterns. First, the point estimates for the short-term effects (2020--21) are consistently positive and highly statistically significant ($p<0.01$) for all three states. Second, the estimated effects are statistically and economically larger for more locked-down states. For instance, the estimated increase in the WFH share in 2020--21 is 11.2 pp in the ACT, 24.1 pp in NSW and 34.9 pp in Victoria. These differences~--- between the estimates for NSW and ACT \emph{and} the estimates for Victoria and NSW~--- are statistically significant at the 1\% level (see Appendix Table~\ref{tab: dose}), highlighting the expected dose--response relationship during the pandemic period. Third, while the medium-term effects are consistently smaller in magnitude, with greater levels of persistence for the extensive margin of working from home and hybrid working arrangements, the pattern is identical: the estimated effects are consistently positive, with the effect sizes increasing with lockdown intensity. For example, the estimated increase in the WFH share in 2022--23 is 2.1 pp in the ACT ($p \geq 0.1$), 8.9 pp in NSW ($p < 0.01$) and 12.3 pp in Victoria ($p < 0.01$). Again,  these differences in the estimates are statistically significant at the 1\% level (see Appendix Table~\ref{tab: dose}), confirming that the strong dose--response relationship persists into the medium term.

This geographic gradient of effect sizes  provides compelling evidence that lockdown intensity matters: stronger, longer shocks produce more durable behavioral change.



\pdfbookmark[1]{Discussions}{section5} 
\section{Discussions}
\label{section5}
As discussed, a substantial portion of the pandemic-period increase in WFH endured into the medium run~--- on average about one-third of the impact of the extended lockdowns on WFH  in treated states persists  through to 2023. In this section, we explore potential mechanisms behind this persistence, drawing on complementary indicators from the employer and employee sides, to examine how their responses may have shifted workplaces in these states into a new equilibrium with higher baseline levels of WFH. 

\pdfbookmark[2]{Responses by Employers}{employers} 
\subsection{Responses by Employers: Office Usage and Hiring}
A key mechanism for WFH persistence is the adaptation by employers and the institutionalization of remote work in standard business practices. Firms in the treated states not only shifted to WFH out of necessity during lockdowns but also have learned from that experience and updated their policies and workplace organization. 

\begin{figure}[htbp]
    \centering
    \caption{Trends in office vacancy rates in  central business districts}
    \label{fig:office_vacancy_rates}
    \includegraphics[width=\textwidth]{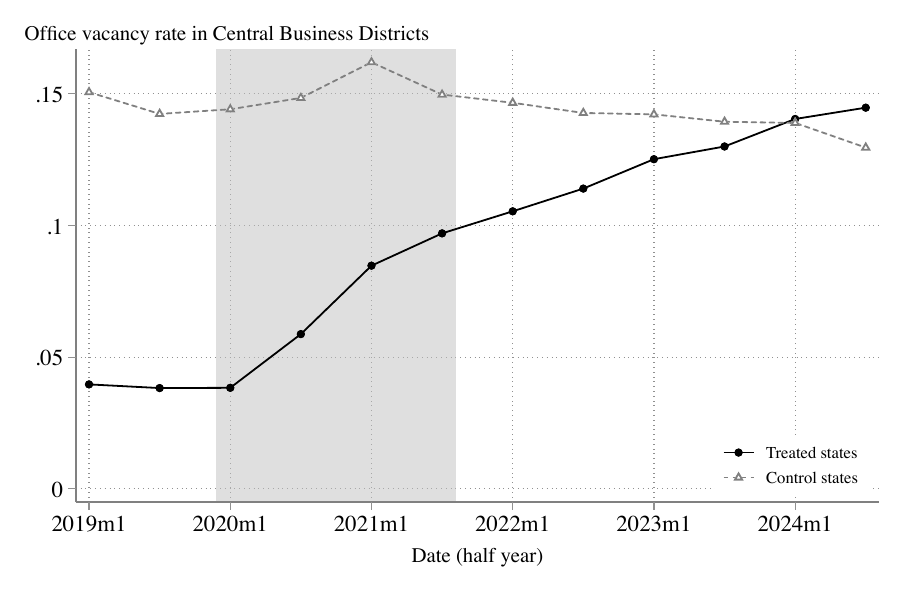}

    \begin{minipage}{\textwidth}
{\footnotesize \underline{Notes}: This figure plots the average office vacancy rates in the central business districts of treated and control regions from 2019 to 2024, using data from the \href{https://www.propertycouncil.com.au/}{Property Council of Australia}. The measure captures the proportion of unoccupied office space in central business districts for each of the capital cities of the six states and two territories. We construct a weighted average for the treated/control regions based on the population size of each city in 2024. The gray shaded area shows the period where lockdowns were in place across parts or all of Australia. See Appendix Figure~\ref{fig:office_vacancy_rates_det} for the rates for individual states.} 
\end{minipage}
\end{figure}

\noindent \textbf{Office Usage.} Pandemic-induced office closures can have both immediate and lasting effects on firms' demand for physical space. Hence, office vacancy rates are a useful indicator of long-term adjustments in workplace practices. In Figure \ref{fig:office_vacancy_rates}, we plot office vacancy rates for the central business districts (CBD) of Australia's state capitals using data published online by the Property Council of Australia. For simplicity, we group the vacancy rates for the three capital cities in treated states and the five control states, weighting each city by 2024 population size. In treated states, vacancy rates have climbed significantly since the second half of 2020, reaching a level 3.5-times the pre-pandemic level. Further disaggregation by individual cities (Appendix Figure~\ref{fig:office_vacancy_rates_det}) shows the largest increase in Melbourne (Victoria) followed by Sydney (NSW), reinforcing our dose--response results discussed above. In contrast, in cities that experience milder lockdowns, office vacancies actually decline from their 2020--21 peaks as normal operations resumed.\footnote{While the pre-COVID levels are very different across treated and control regions, the trends are similar from 2019 to the middle of 2020.} The growing surplus of office space in Sydney and Melbourne implies that many firms in those areas do not fully revert to pre-pandemic on-site work. This pattern is consistent with employers internalizing the lessons of the lockdown: when investments in remote-working infrastructure are made and new routines are established, the marginal benefit of maintaining large offices falls. This kind of organizational change can be self-reinforcing. 

\noindent \textbf{Job Postings allowing WFH.} Another clear sign of employers' responses is the formalization of remote work arrangements. As discussed in Table \ref{tab: did}, workers in treated regions are 21.3 pp and 10.2 pp more likely to have a formal WFH agreement with their employer during and after the pandemic. 

\begin{figure}[htbp]
    \centering
    \caption{Share of office-job ads in major urban areas allowing WFH over time by state}
    \label{fig:job_postings}
    \begin{subfigure}[t]{0.8\textwidth}
    \caption{Raw data}
    \includegraphics[width=\textwidth]{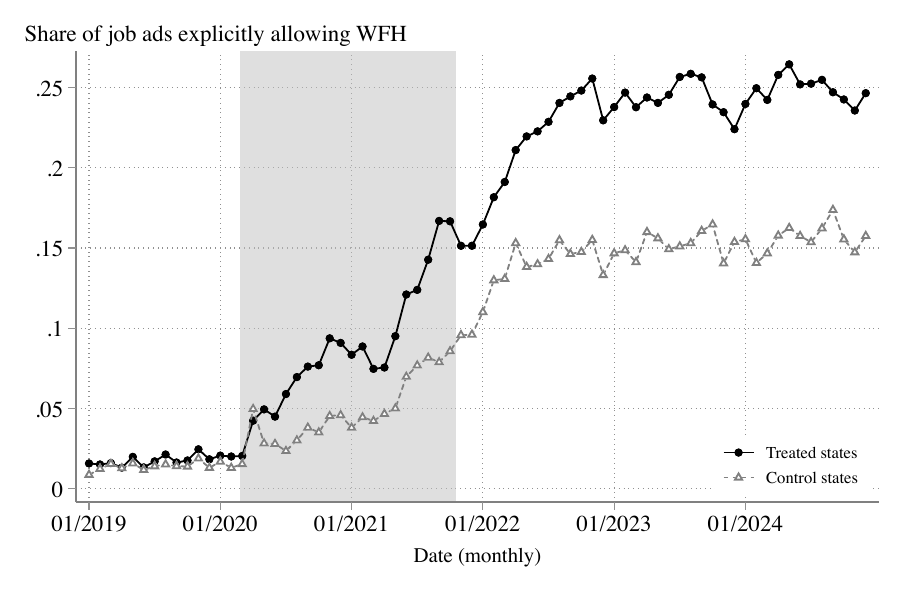}  
    \end{subfigure}
    \begin{subfigure}[t]{0.8\textwidth}
    \caption{Dynamic DiD estimates, with occupation-month fixed effects and other controls}
    \includegraphics[width=\textwidth]{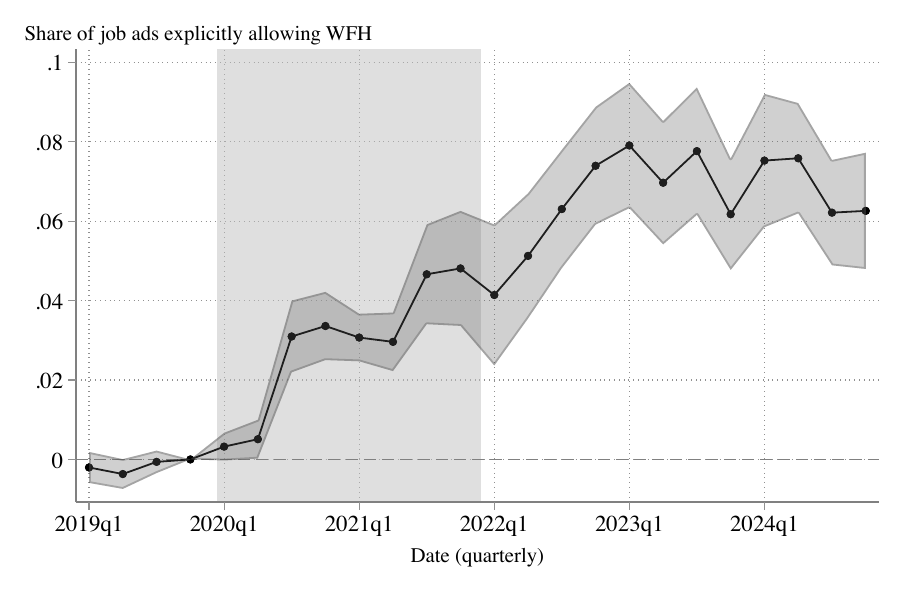}  
    \end{subfigure}

 \begin{minipage}{\textwidth}
{\footnotesize \underline{Notes}: This figure plots the share of job advertisements that explicitly allow WFH arrangements from 2019 to 2024. The data, provided by \cite{hansen2023remote}, uses natural language processing to extract WFH-related language from the near-universe of online job ads in English-speaking countries. We obtained custom data from the authors for the whole of Australia at the subregion--state--month--year--occupation--industry level. To match our HILDA sample, we restrict to office job occupations in major urban areas. In Panel (b), we present the point estimates and 95\% confidence intervals from a dynamic difference-in-differences regression similar to Specification~\eqref{eq:eventstudy}. We weight cells by the total number of job ads and cluster standard errors at the state-by-occupation level.} 
\end{minipage}
\end{figure}

Consistent with previous findings, new job postings increasingly begin to highlight WFH options. Figure \ref{fig:job_postings}, drawing on data from \citet{hansen2023remote} based on the text of job vacancy postings, plots the share of office-job ads explicitly offering WFH at least one day per week each month from 2019 to 2024. Prior to the pandemic, only about 2\% of postings include such language. This share rises sharply in 2020 for both treated and control regions and remains elevated through 2023. By early 2023, nearly one-quarter of new office job ads in the treated regions include WFH or hybrid options, compared to roughly 15\% in control states that experienced milder lockdowns. The level of WFH postings has since stabilized, with treated states maintaining a consistent 6--8 pp gap, even after we account for differences in occupations and industries across states.\footnote{In panel (b), we estimate a dynamic DiD regression similar to Specification 4.1, with industry and occupation by month fixed effects, and standard errors clustered by state and occupation. The unit of observation is the share of job posts in a given month, state, occupation and industry. We weight our estimates by the total number of job posts in each cell.} This estimate is similar to our estimated post-COVID effect of 10.2 pp in the HILDA data (which is based on both existing jobs and new jobs). The evolution in WFH job ads suggests that WFH arrangements have likely persisted at a relatively stable level beyond 2023, when our HILDA sample ends.\footnote{While the post-COVID effect using the job postings data is similar to our estimate (10.2 pp compared to 6--8 pp), the patterns during the lockdown period differ. Unlike our HILDA-based measures, Figure \ref{fig:job_postings} shows a milder increase in WFH during the lockdown. We interpret this discrepancy as reflecting that, during strict lockdowns, remote work for office jobs was often implicitly assumed and therefore not explicitly mentioned in job ads.} This sizable difference in job-posting trends indicates that employers in treated regions become far more open to hiring for remote roles, reflecting an institutional change in work norms. These patterns are consistent with developments in other countries, including the US, the UK, and New Zealand.

In sum, employer actions in the treated states~--- from real estate decisions to HR policies~--- reveal a conscious adaptation that has locked in a higher baseline level of WFH. The temporary shock creates an institutional path-dependence: having invested in remote work systems and seen their benefits, firms have incorporated them into standard practice. 

\pdfbookmark[2]{Responses from Workers}{workers} 
\subsection{Responses from Workers: WFH Investments and Relocation}
On the worker side, prolonged exposure to WFH has potentially led to lasting changes in preferences and habits that underpin the persistence of remote work. Employees in the treated states experience extended periods of mandatory home-working, allowing them to acclimate to this new mode of work. 

\begin{figure}[htbp]
    \centering
    \caption{Examining adoption of home-office technologies/infrastructure via Google Trends data}
    \label{fig:google_trends}
    \begin{subfigure}[t]{0.49\textwidth} \caption{``Computer''}\includegraphics[width=\textwidth]{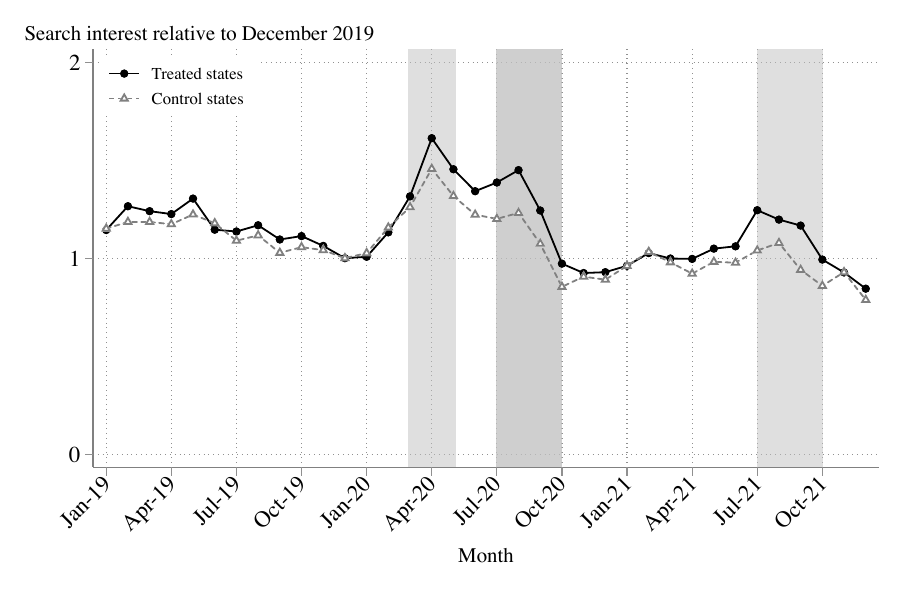}   
    \end{subfigure}
    \begin{subfigure}[t]{0.49\textwidth}
    \caption{``Computer monitor''}\includegraphics[width=\textwidth]{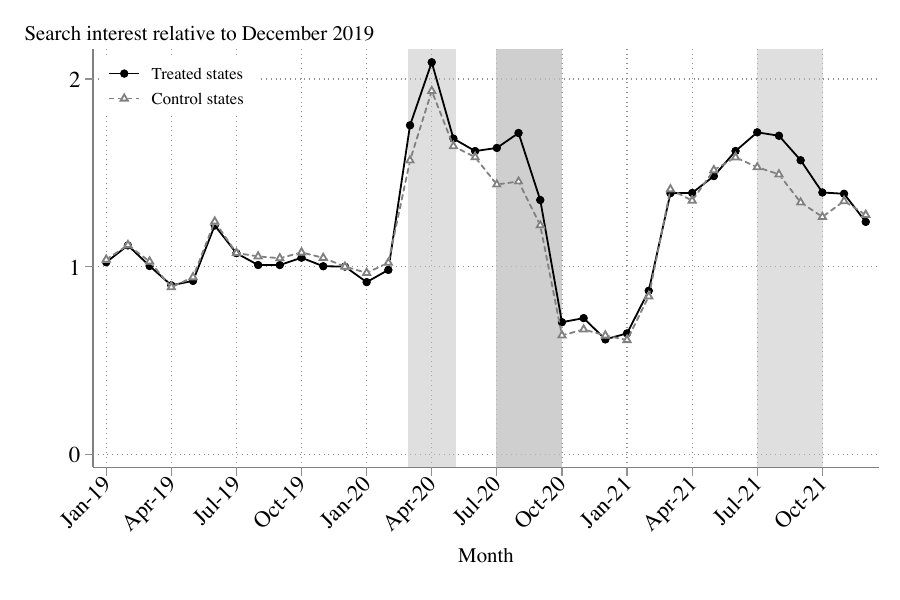}   
    \end{subfigure} \vspace{0.1cm}

       \begin{subfigure}[t]{0.49\textwidth} \caption{``Desk''}\includegraphics[width=\textwidth]{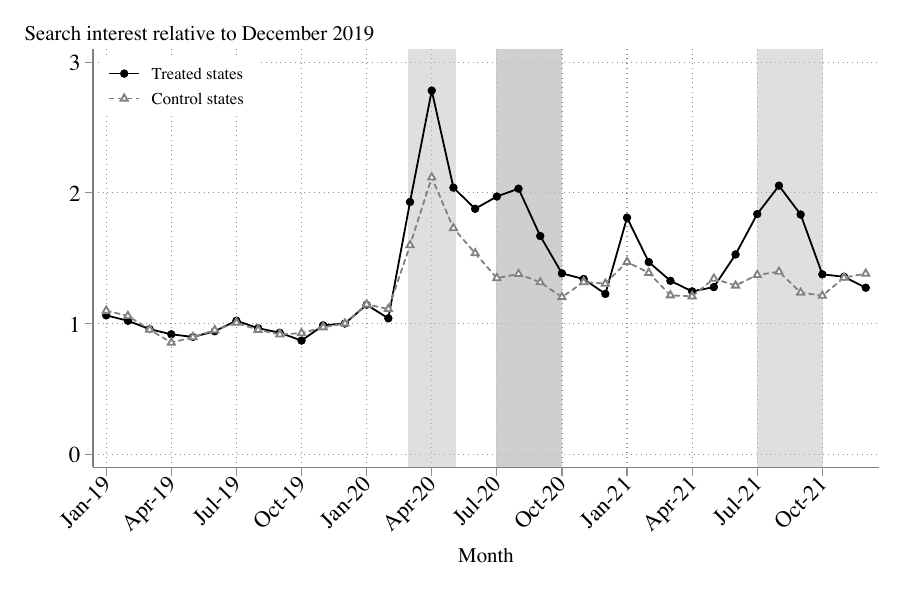}   
    \end{subfigure}
    \begin{subfigure}[t]{0.49\textwidth}
    \caption{``Office chair''}\includegraphics[width=\textwidth]{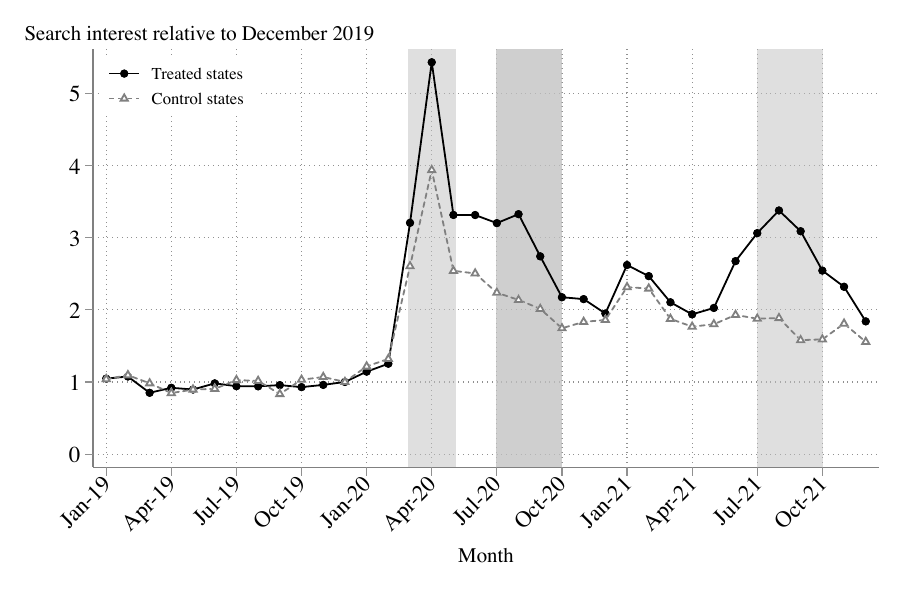}   
    \end{subfigure} \vspace{0.1cm}

   \begin{subfigure}[t]{0.49\textwidth} \caption{``Zoom''}\includegraphics[width=\textwidth]{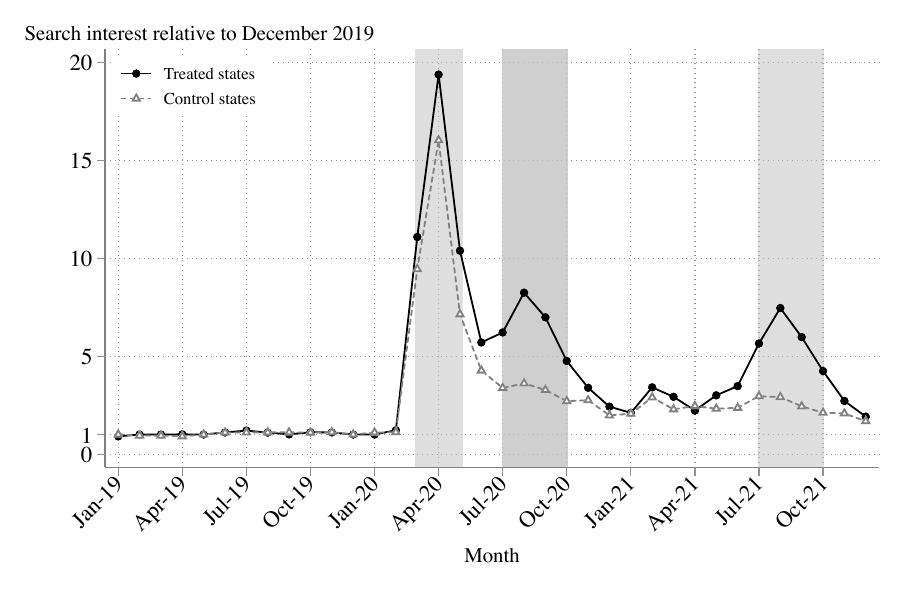}   
    \end{subfigure}
    \begin{subfigure}[t]{0.49\textwidth}
    \caption{``Microsoft Teams''}\includegraphics[width=\textwidth]{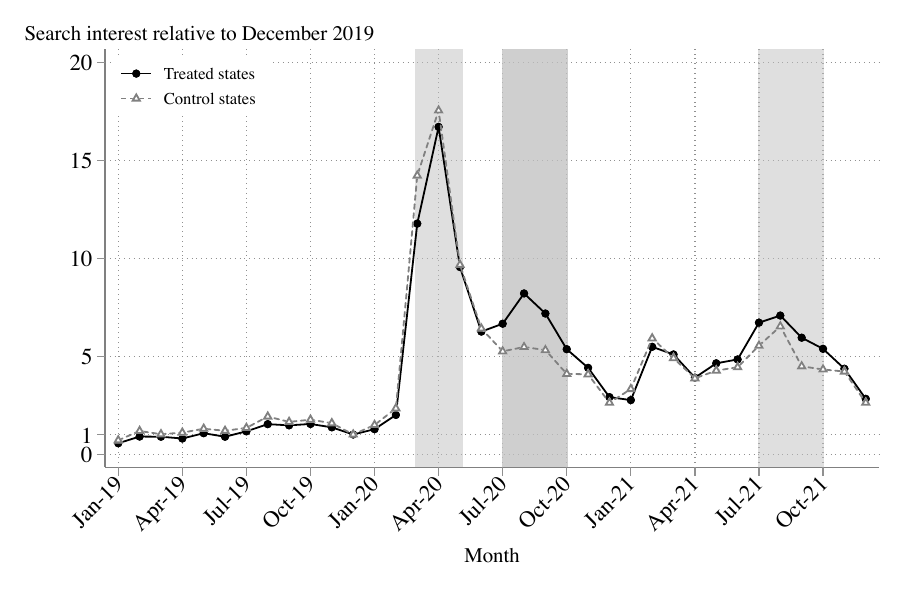}   
    \end{subfigure} 

 \begin{minipage}{\textwidth}
{\footnotesize \underline{Notes}: These figures plot relative search interest for each specified keyword from 2019 to 2021, using data from \hyperref[https://trends.google.com/trends/]{Google Trends}. We weight the state-level monthly indices from Google Trends by the population size of the largest city in each state, to construct a weighted average for treated and control states over time. We then normalize our indices by the relative amount of search interest in treated/control regions in December 2019.} 
\end{minipage}
\end{figure}

\noindent \textbf{WFH Investments.} In response to the stricter and longer lockdown requirements, workers may have responded by investing in home office technology and skills, which facilitated effective remote work and thus encouraged its continuation. Data from Google search behavior provide revealing insights into this process. Figure \ref{fig:google_trends} tracks Google Trends indices for various WFH-related technologies and equipment. It shows a surge in searches for terms like ``computer monitor" and ``office chair", and teleconferencing software (e.g., ``Zoom") in all states, during the national lockdown. Importantly, interests in these tools show clear spikes in treated states during the Melbourne lockdown period and the Delta-strain lockdowns, and these periods do not coincide with spikes in the control states. Not surprisingly, the spikes in treated states are driven respectively by Victorians in mid-2020 and NSW/ACT in mid-2021 (Figure~\ref{fig:google_trends_app2}). Although we cannot directly measure the purchase of office furniture or software downloads, these trends suggest that extended lockdowns in treated states may have caused many workers to invest in more comfortable and productive home offices and embrace online meeting technologies. These investments have a long-lasting payoff: once a worker has set up a proper home office and mastered remote collaboration tools, the incremental cost (in terms of effort or lost productivity) of continuing to WFH is much lower. 

\noindent \textbf{Relocation.} Prolonged lockdown may also make employees relocate further from current/future employers and therefore reduce their incentives to go back to the office. In Table \ref{tab: mover} columns (1) to (3), we use our HILDA analysis sample to examine whether office workers in treated regions are more likely to change residence during the lockdown period (2020–21) or afterward (2022–23); workers that move states are excluded from the analysis. During the lockdown period, employees in treated regions are not more likely to move. By 2022--23, however, a notable gap emerges: office workers in treated states become 3.1 pp (16\%) more likely to have moved than those in control states. Somewhat surprisingly, this post-lockdown relocation effect is driven almost entirely by homeowners in the treated regions. Homeowners under prolonged lockdowns show a 3.1 pp (29\%) higher probability of moving after the pandemic (significant at the 5\% level). In contrast, we observe a near-zero DiD coefficient for renters in both periods, suggesting that there is essentially no differential change in the treated regions.

Next, we want to investigate whether employees are relocating farther away from their workplace, which can potentially reinforce their persistence of WFH after lockdowns. Since the reported home-to-work distance in the HILDA survey is confounded by workers' WFH choices,\footnote{The survey question asks about the person's usual place of work not the location of their employer.} we instead use the distance from a mover's new residential postcode to the CBD of the state capital. This allows us to assess whether relocation patterns reflect a shift toward living farther from city centers, where a high proportion of offices are located.

During the immediate lockdown period, workers in treated regions who moved tend to settle 28.0 km farther from the CBD, a gap that widens to 37.7 km post-COVID, representing increases of 33\% and 45\%, respectively. Homeowners show a delayed but substantial response: while there is no significant difference in distance moved during the lockdown, by 2022--23, treated-region homeowners are relocating 60.3 km farther from the CBD, a 89\% increase relative to their counterparts in control regions. Renters, who are generally more mobile, respond more quickly: in 2020–21, renters in treated regions move 39.3 km farther than those in control regions, a 42\% increase; and this effect persists in 2022–23 at 43.1 km (46\%). Notably, this shift occurs even though the overall likelihood of moving among renters does not increase. 

These patterns suggest that prolonged lockdowns prompted lasting moves to more distant areas, likely in outer suburbs or satellite cities within commuting distance of Sydney or Melbourne, such as Newcastle, NSW or Geelong, Victoria.\footnote{Since our sample focuses on workers in major urban areas (with 100,000 or more residents), moves to smaller regional areas will not be captured.} Such relocation responses reduce workers' flexibility to return to centralized workplaces once restrictions are lifted. Moreover, as a significant share of the shift reflects homeowners purchasing new residences farther from city centers~--- and with high transaction costs on property purchases~--- these relocation effects will likely reinforce a continued reliance on WFH in treated regions going forward.

\pdfbookmark[1]{Robustness and Heterogeneity}{section6} 
\section{Robustness and Heterogeneity}
\label{section6}
In this section, we first address potential threats to our identification and then explore heterogeneity in WFH uptakes across occupations and demographic groups.

\pdfbookmark[2]{Robustness}{robustness} 
\subsection{Robustness} 
\label{robustness}
\textbf{Pre-existing Regional Variations.} One potential concern for our identification is that Victoria, NSW, and the ACT ~--- the three treated regions that experienced the most prolonged lockdowns ~--- may have structural characteristics that would lead to higher WFH adoption after the national lockdown, even in the absence of the additional region-specific lockdown restrictions. These regions include Australia's two largest metropolitan areas (Sydney and Melbourne), which tend to have a higher concentration of WFH-suitable occupations, larger populations, longer average commute times, greater urban density, and more expensive CBD office costs, all of which could contribute to higher WFH uptake after it became normalized during the pandemic regardless of lockdown intensity.

We present several pieces of evidence that help mitigate this concern. First, our estimates with period-specific controls (Table~\ref{tab: did}, Panel B) account for many of these factors, including occupational and industry differences, and pre-COVID postcode-level average commuting times among full-time workers. The fact that these controls only marginally reduce the estimated effects suggests that pre-existing differences across states is unlikely to explain a major fraction of our results. Besides, the inclusion of state-by-occupation fixed effects allows for the possibility that certain occupations (e.g., clerks) may be more amenable to remote work in larger, more urbanized states like Victoria than in less densely populated states, addressing the concern from confounding structural characteristics that might bias our estimates. 

\begin{figure}[htbp]
\centering
\caption{Dose--response estimates: Heterogeneity by state / additional days in lockdown} \label{fig: dose}

\begin{subfigure}[t]{0.49\textwidth}
\caption{WFH share}
\includegraphics[width=\textwidth]{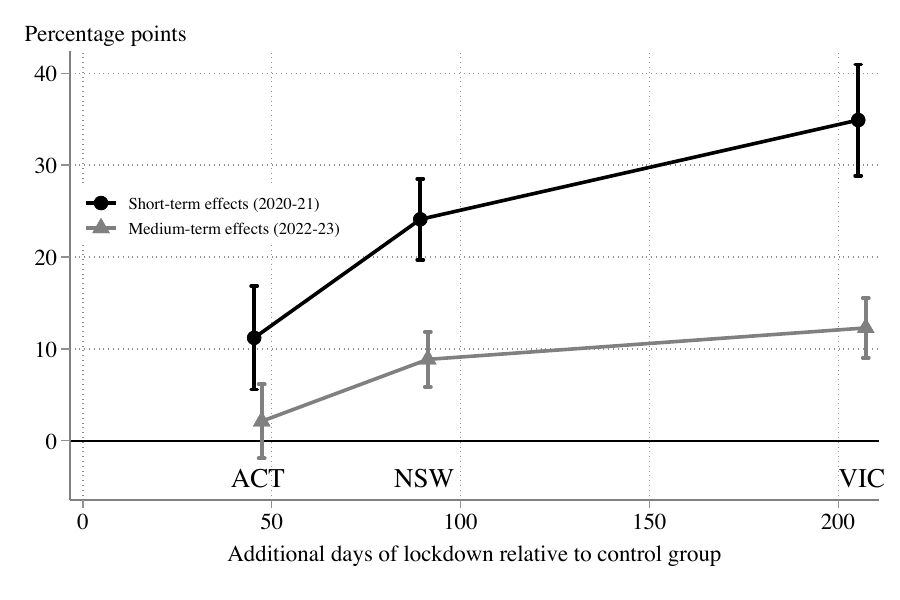}
\end{subfigure}
\begin{subfigure}[t]{0.49\textwidth}
\caption{WFH hours per week}
\includegraphics[width=\textwidth]{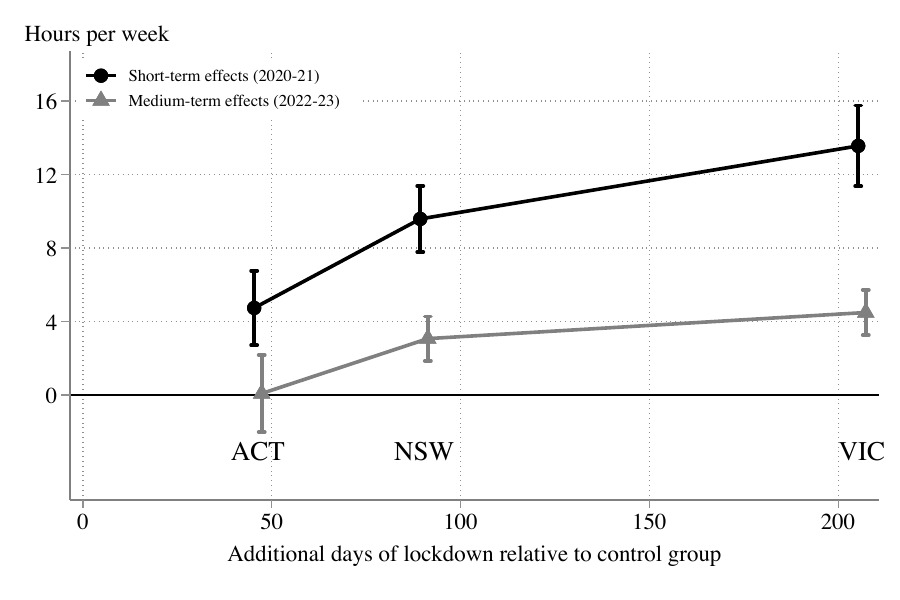} 
\end{subfigure} \vspace{0.1cm}

\begin{subfigure}[t]{0.49\textwidth}
\caption{Any WFH}
\includegraphics[width=\textwidth]{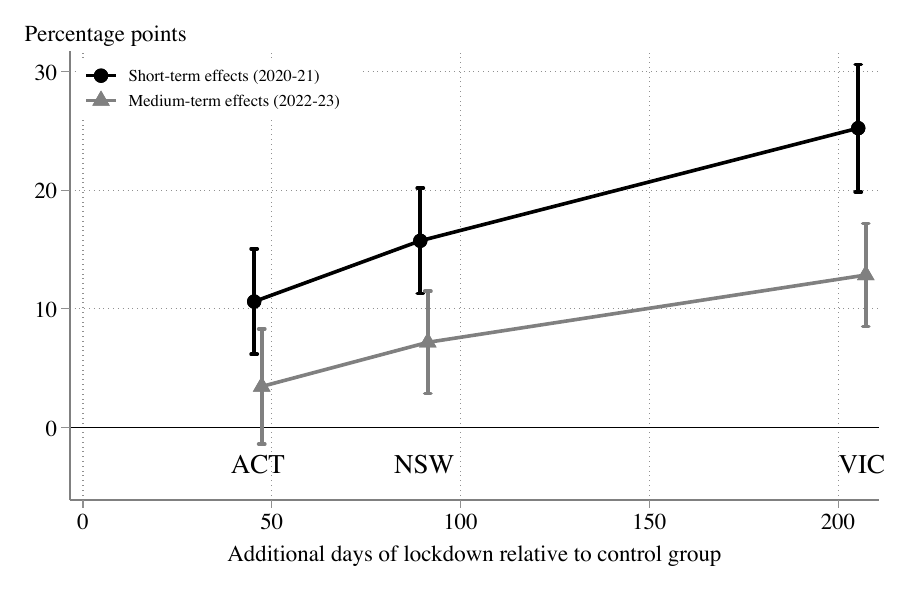} 
\end{subfigure}
\begin{subfigure}[t]{0.49\textwidth}
\caption{Any WFH with formal agreement}
\includegraphics[width=\textwidth]{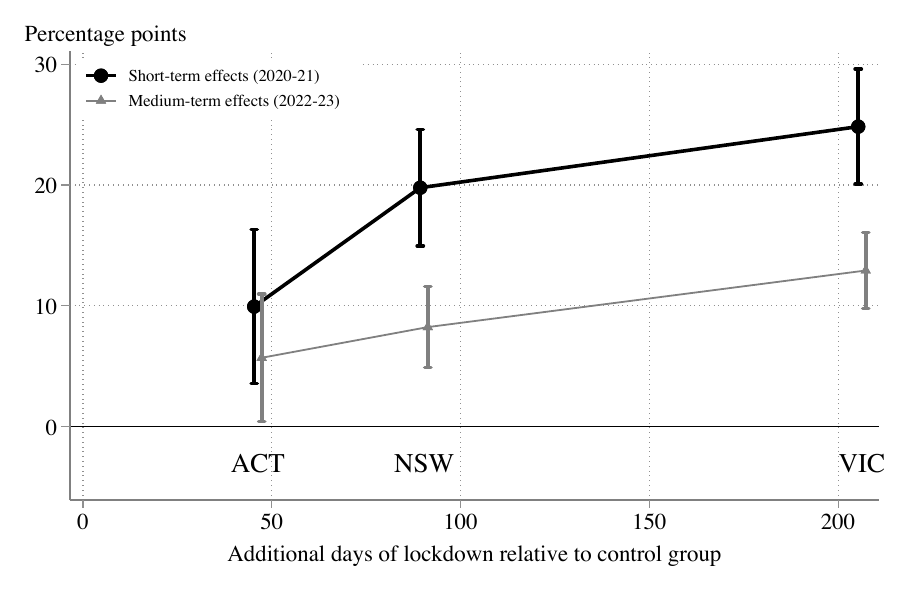} 
\end{subfigure} \vspace{0.1cm}

\begin{subfigure}[t]{0.49\textwidth}
\caption{Mostly WFH (WFH share $\geq$ 0.6)}
\includegraphics[width=\textwidth]{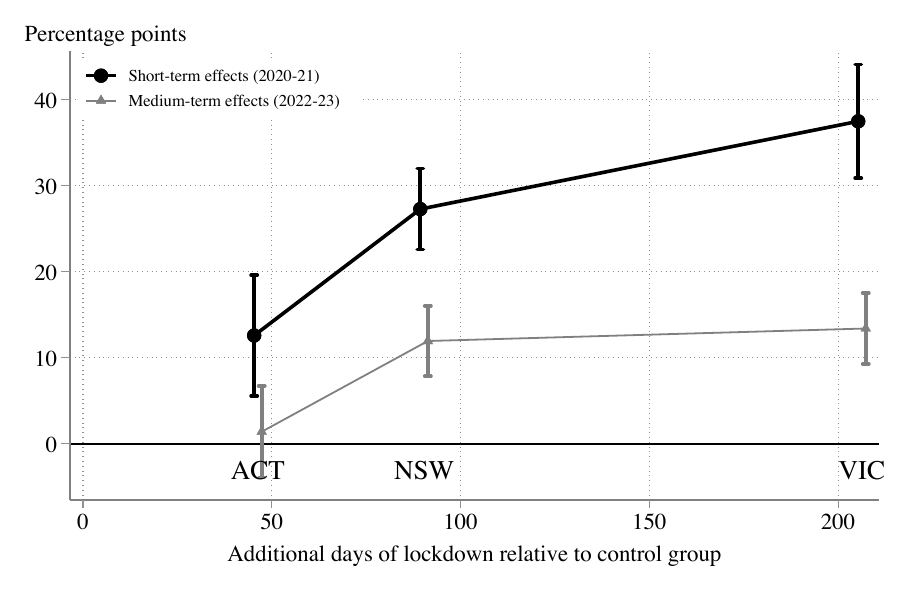}
\end{subfigure}
\begin{subfigure}[t]{0.49\textwidth}
\caption{Fully remote (WFH share $=$ 1)}
\includegraphics[width=\textwidth]{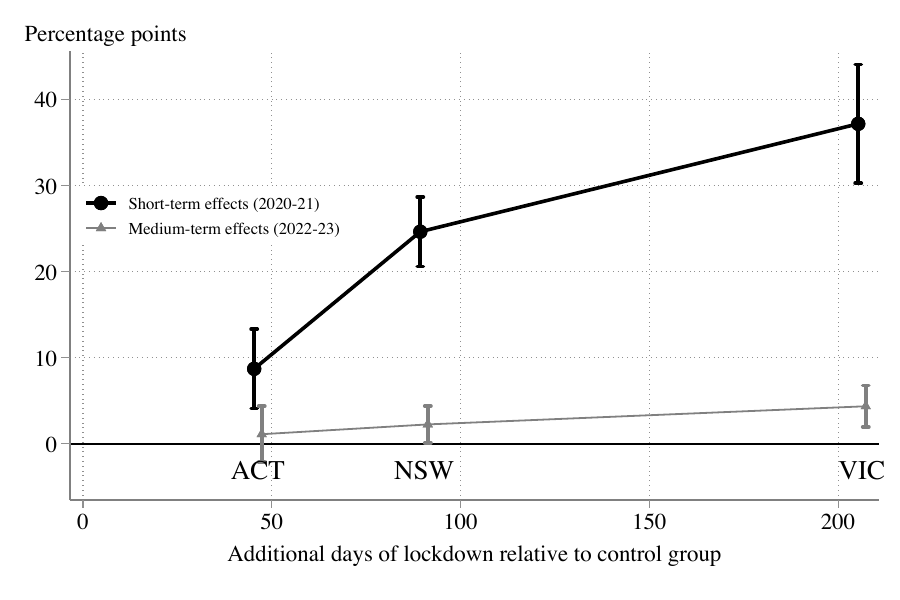} 
\end{subfigure}

\begin{minipage}{\textwidth}
{\footnotesize \underline{Notes}: These figures plot the estimated treatment effects for each of the three treated regions, based on Specification \ref{eq:basicDiD}, against the number of additional lockdown days relative to the control regions (shown on the x-axis). The corresponding coefficient estimates are reported in Table \ref{tab: dose}.
 } 
\end{minipage}
\end{figure}

Second, among the treated regions, Sydney (NSW) has a larger office workforce, greater population, and higher office costs than Melbourne (Victoria). If these structural factors were the primary drivers of WFH adoption, we would expect larger effects in NSW. However, as shown in Figure~\ref{fig: dose}, we find the opposite: Victoria exhibits larger and more persistent WFH effects, consistent with its longer and more stringent lockdown duration. This pattern strongly suggests that policy exposure, rather than pre-existing structural differences, is driving the differential WFH outcomes.

Third, we implement additional robustness checks to address the above concerns, presented in Figure \ref{fig: robust}. In one exercise, we restrict the sample to postcodes whose pre-COVID average estimated commute times are comparable to those in treated regions; and in another practice, we limit the sample to capital cities across states, ensuring that both treated and control groups consist of highly urbanized areas. In both cases, we continue to find consistent treatment effects, reassuring us that the estimates genuinely reflect responses to variation in lockdown intensity.

\noindent \textbf{COVID Infections and Worse Health Conditions.} Another possible explanation of our observed effect is that more severe COVID outbreaks in the treated regions may have adversely affected workers' health, making them persistently more inclined or more suited to work from home. For instance, higher infection rates or heightened health anxiety could limit their ability or willingness to return to the office. Overall, we find mixed evidence: while workers in treated regions report higher levels of COVID-related anxiety and slightly greater infection exposure, we observe little indication of worsened general health, reduced work capacity, or lasting behavioral changes that would substantively account for the persistent WFH patterns.

Panel A of Table \ref{tab: anxiety} presents the post-pandemic COVID-related variables. Columns (1)--(4) examine COVID-related anxiety and crowd discomfort, surveyed only in 2022--23; Columns (5)--(6) use 2022 responses on COVID infections. Since these variables are only surveyed after the pandemic's acute phase, we directly compare treated and control regions.\footnote{We exclude state-by-occupation fixed effects in Panel A, since these terms are collinear with treatment, but include all other controls.} We find that workers in treated regions report higher levels of anxiety and discomfort in crowds, which may contribute to a greater preference for WFH. However, recent infection rates are similar across regions, and cumulative infections are only slightly higher in treated states. Moreover, most infections occurred among vaccinated individuals, as restrictions were relaxed and the highly infectious but less virulent Omicron-variant emerged in late 2021. Thus, long-term health differences resulting from COVID-19 infections are likely to be relatively minor between workers in treated and control states.


To complement the limited indicators in Panel A, we examine a broader set of health-related variables in Panel B to assess whether workers in treated regions experience any decline in health that may hinder their return to office.\footnote{Panels B and C replicate the main DiD specification with period-specific controls.} We consider self-rated general health, physical and mental difficulties performing work, and the prevalence of long-term health conditions; across these measures, we find little evidence of worse health outcomes in treated regions.\footnote{We use self-assessed health (1 = excellent, 5 = poor) in Column 1; Columns 2–4 use indicators for physical difficulty, mental difficulty, or long-term health conditions; column 5 uses an indicator for shortness of breath, which is surveyed conditional on having a long-term condition.} In fact, the incidence of long-term health conditions is actually 2.8 pp (significant at 10\% level) lower in treated regions, a 21\% reduction relative to the pre-COVID mean of 13.2\%, and we observe no increase in symptoms such as shortness of breath. Additionally, we find no significant difference in reported weekly working hours, further suggesting that workers' capacity to work is not substantially impaired.

Panel C explores household spending responses that may reflect health-related needs, health concerns and related behavioral changes.\footnote{We use self-reported household expenditure data from HILDA, where all measures are winsorized at the top 99th percentile. Spending on public transit and fuel is adjusted for working hours in Columns (5) and (6).} Columns (1) and (2) focus on health-related expenditures. Somewhat surprisingly, households in treated regions report significantly lower annual spending on health practitioners and medications, both during and after the pandemic. While we cannot determine whether these expenses are directly tied to COVID-related conditions, the consistent decline in health spending does not support concerns of widespread health deterioration in treated regions.

Regarding behavioral indicators, we find little support for the idea that workers in treated regions exhibit heightened health concerns or anxiety that can drive persistent WFH. Interestingly, households in treated states spend significantly more on cigarettes or tobacco~--- up 39\% during the lockdown and 31\% post-COVID~--- despite the well-known elevated health risks of smoking during the pandemic. This pattern suggests that health-related concerns are not stronger in treated regions, casting doubt on the argument that persistent WFH in these regions is driven by an ongoing fear of infection. For other indicators, we observe that treated-region workers spend less on meals eaten out and on transportation (both public transit and fuel) during the lockdown period. However, these reductions dissipate by 2022--23. The quick rebound in dining and travel behaviors does not suggest sustained anxiety or health-related avoidance of in-person settings. 

Taken together, the results from Panels B and C suggest that although workers in treated regions experienced more anxiety and slightly higher infection rates, their physical health, ability to work, and long-run behavioral patterns do not appear to have changed enough to explain the persistent WFH effect. Thus, while health-related factors may play a role at the margin, the evidence does not support the interpretation that health deterioration was a major driver of continued WFH in treated states.\footnote{Our results are also robust to controlling for a range of health-related factors ~--- including COVID-related anxiety and infection history (coded as zeros for pre-COVID observations to enable DiD estimation), long-term health conditions, and self-assessed health status ~--- with our main coefficients of interest remaining largely unchanged.}

\noindent \textbf{More Robustness: Alternative Estimations.} We conduct a range of robustness exercises using alternative fixed effects, model specifications, inference and clustering methods, treatment definitions, and sample restrictions. As expected, the estimated effects are somewhat attenuated when we expand the sample to include non-office jobs or workers in smaller urban areas; in all the other exercises, our main findings remain highly similar in magnitude to our baseline estimates. We present and discuss these robustness checks in more detail in Appendix \ref{more robustness}.

\pdfbookmark[2]{Heterogeneity}{heterogeneity} 
\subsection{Heterogeneity}\label{heterogeneity}
\noindent \textbf{Occupations' Pre-COVID WFH Level.} Even within our broad office-worker category, occupations vary in the nature of tasks and the feasibility of remote work. Figure \ref{fig: heterogeneity} highlights the relationship between an occupation's WFH share before the pandemic, its WFH uptake during the pandemic, and its WFH persistence after the pandemic. Each point in the figure represents a major occupation group, with blue circles for office jobs and red triangles for non-office jobs (included here for comparison). 

Figure~\ref{fig: heterogeneity}(a) plots the change in WFH share during the lockdown period (2020–21) against the change in WFH share in the post-pandemic period (2022–23) for each occupation. There is a clear positive relationship: occupations that experienced larger increases in WFH during the lockdown tended to retain higher levels of WFH later on. In other words, the more an occupation was able to adapt to remote work out of necessity, the more it continued to utilize remote work after the immediate necessity passed. Office jobs cluster in the upper-right portion of the plot, showing both big initial jumps and substantial persistence, whereas non-office occupations are near the origin, with tiny WFH increases and almost no lasting change post-pandemic. 

Figure~\ref{fig: heterogeneity}(b) examines why certain occupations saw more WFH uptake during the lockdown, by plotting the lockdown-period WFH increase against the pre-COVID WFH share for each occupation. Occupations that already had some prevalence of WFH before 2020 were the ones that could most readily expand WFH during the pandemic. These occupations~--- primarily white-collar professional jobs~--- had the necessary work processes in place, or at least the potential to develop them, to support remote work. By contrast, occupations with essentially zero pre-pandemic WFH, mostly non-office jobs, could only temporarily shift a small fraction of tasks online and therefore saw minimal lockdown-induced WFH increases. 

Lastly, Figure~\ref{fig: heterogeneity}(c) examines the relationship between an occupation's pre-COVID WFH share and its post-COVID WFH persistence. For non-office jobs, WFH levels largely reverted to pre-pandemic norms once lockdowns ended, as indicated by the near-flat line at zero. In contrast, office jobs were more likely to sustain elevated WFH levels. However, the much flatter slope compared to the lockdown period in  Figure~\ref{fig: heterogeneity}(b) suggests that the pandemic fundamentally altered the relationship between pre-existing WFH patterns and subsequent uptake. In particular, medium-run WFH persistence becomes less strongly tied to pre-COVID levels, implying that the lockdown shock changed norms and practices in ways that made post-COVID WFH behavior deviate from its historical baseline.

This heterogeneity by occupation indicates that the feasibility of remote work is a key determinant of its persistence, consistent with prior research highlighting large variability in WFH possibilities across different jobs \citep{adams2022work, bartik2020jobs}.

\noindent \textbf{Heterogeneity by Employee Characteristics.} Next, in Figure \ref{fig: correlation}, we plot the correlation between the immediate WFH effect during the 2020-21 lockdown and the medium-run WFH effect observed in 2022-23, across demographic groups. It reveals a clear positive correlation: subgroups that experienced larger lockdown-induced increases in WFH~--- such as university-educated workers, women, and mid-career workers~--- also tend to retain higher WFH rates post-pandemic. Somewhat surprisingly, parents of young children do not sustain higher WFH levels, likely reflecting a universal lockdown shift and the easing of childcare burdens post-COVID. We present more detailed analysis of heterogeneity in Appendix~\ref{more heterogeneity}.

\pdfbookmark[1]{Conclusion}{section8} 
\section{Conclusion} \label{section8}
This paper examines the persistence of WFH practices in the aftermath of the COVID-19 pandemic, focusing on office workers in Australia. Leveraging the stark differences in lockdown intensity across Australian states and territories, we estimate a series of difference-in-differences models to assess the immediate and medium-term impacts of COVID-19 exposure on WFH adoption. We find strong evidence that the duration and intensity of COVID-19 lockdowns have a significant impact on the adoption and persistence of remote work, with WFH levels remaining elevated two years after the most stringent lockdowns had ended. Approximately 11--49\% of the lockdown-induced increase in WFH is still evident by 2022–23, suggesting that the temporary government restrictions have triggered lasting changes in work habits rather than a transitory spike in remote work. Overall, the evidence indicates that the government policies have significantly contributed to ``a new equilibrium" in which WFH arrangements and hybrid models become an enduring feature of the post-pandemic labor market.

This study complements prior cross-country research by providing the first quasi-experimental evidence of lockdowns' impact on remote work adoption. The results confirm that more stringent and prolonged lockdowns yield significantly more persistent increases in WFH, aligning with those global patterns, while firmly establishing a direct link by isolating the lockdown effect in a relatively homogeneous context.

This persistent shift carries broad implications for labor market dynamics and policy, demonstrating how a temporary health shock can lead to lasting, uneven changes across regional labor markets. The adjustments made by both employers and employees suggest that cities in the hardest-hit states may continue to experience reduced demand for downtown office space, along with increased residential movement away from urban centers, while less-affected regions may see a more complete return to pre-pandemic norms. Over time, these divergent trajectories in work practices could contribute to the emergence of ``a new equilibrium" that extends beyond the labor market and reshapes the urban economy in fundamental ways.

\clearpage
\pdfbookmark[1]{References}{bib} 
{
\setstretch{1}
\footnotesize
\bibliographystyle{aea}
\bibliography{wfh}
}

\clearpage
\pdfbookmark[1]{Figures and Tables}{figures} 

\clearpage
\clearpage

\clearpage
\clearpage 

\clearpage 
\clearpage

\clearpage
\pagebreak

\clearpage

\clearpage

\clearpage

\clearpage


\clearpage
\begin{table}[htbp]
\small
\begin{threeparttable}
\caption{Short and medium-term effects of stronger pandemic-exposure on WFH measures} \label{tab: did}
\begin{tabular}{lcccccc}              \hline \hline & WFH & WFH & Any & WFH with & WFH & WFH \\ & share & hours & WFH & agreement & share $\geq 0.6$ & share $=1$ \\ & (1) & (2) & (3) & (4) & (5) & (6) \\ \hline \addlinespace \multicolumn{7}{c}{Panel A: No controls}             \\ \addlinespace Treated $\times$ 2020--21 & 0.313*** & 12.5*** & 0.217*** & 0.240*** & 0.343*** & 0.324*** \\ \addlinespace[0.3em] & (0.043) & (1.7) & (0.038) & (0.039) & (0.045) & (0.041) \\ \addlinespace[0.6em] Treated $\times$ 2022--23 & 0.117*** & 4.3*** & 0.117*** & 0.122*** & 0.130*** & 0.038*** \\ \addlinespace[0.3em] & (0.027) & (1.1) & (0.038) & (0.029) & (0.028) & (0.014) \\ \addlinespace[0.6em] \quad Implied persistence & 37\% & 34\% & 54\% & 51\% & 38\% & 12\% \\ \addlinespace[0.6em] Mean DV & 0.234 & 8.7 & 0.429 & 0.312 & 0.188 & 0.128 \\ \addlinespace[0.3em] R-squared & 0.284 & 0.297 & 0.151 & 0.180 & 0.235 & 0.211 \\ \addlinespace[0.3em] Observations & 57,896 & 57,896 & 57,896 & 57,896 & 57,896 & 57,896 \\ \addlinespace[0.3em] \hline \addlinespace \multicolumn{7}{c}{Panel B: Period-specific controls}             \\ \addlinespace Treated $\times$ 2020--21 & 0.280*** & 11.0*** & 0.196*** & 0.213*** & 0.308*** & 0.291*** \\ \addlinespace[0.3em] & (0.031) & (1.2) & (0.029) & (0.024) & (0.032) & (0.034) \\ \addlinespace[0.6em] Treated $\times$ 2022--23 & 0.100*** & 3.5*** & 0.095*** & 0.102*** & 0.119*** & 0.031*** \\ \addlinespace[0.3em] & (0.016) & (0.6) & (0.021) & (0.017) & (0.020) & (0.009) \\ \addlinespace[0.6em] \quad Implied persistence & 36\% & 32\% & 49\% & 48\% & 39\% & 11\% \\ \addlinespace[0.6em] Mean DV & 0.234 & 8.7 & 0.429 & 0.312 & 0.188 & 0.128 \\ \addlinespace[0.3em] R-squared & 0.431 & 0.448 & 0.263 & 0.284 & 0.367 & 0.322 \\ \addlinespace[0.3em] Observations & 57,896 & 57,896 & 57,896 & 57,896 & 57,896 & 57,896 \\ \hline \end{tabular}

\begin{tablenotes}
\footnotesize
\item *** $p<0.01$, ** $p<0.05$, * $p<0.01$. This table presents the difference-in-differences estimates based on Specification \eqref{eq:basicDiD}. Panels A and B report results without and with period-specific controls $X_{iost}$, respectively. All regressions include state-by-occupation fixed effects and survey wave fixed effects. Standard errors are clustered at the state-by-occupation level. Implied persistence is calculated as the ratio of the post-COVID effect to the lockdown-period effect. The mean of the dependent variable is based on the average value in control states in 2020.
\end{tablenotes}
\end{threeparttable}
\end{table}

\clearpage
\begin{table}[htbp]
\small
\begin{threeparttable}
\caption{Employees' response: Moved house   } \label{tab: mover}
\begin{tabular}{lcccccc}              \hline \hline &\multicolumn{3}{c}{Moved House}      &\multicolumn{3}{c}{Distance to CBD (Movers)}     \\ \cmidrule(lr){2-4}  \cmidrule(lr){5-7} & All & Homeowners & Renters & All & Homeowners & Renters \\ & (1) & (2) & (3) & (4) & (5) & (6) \\ \hline\addlinespace Treated $\times$ 2020--21 & 0.017 & 0.013 & 0.010 & 28.0*** & 4.7 & 39.3** \\ \addlinespace[0.3em] & (0.010) & (0.012) & (0.026) & (9.3) & (23.8) & (16.2) \\ \addlinespace[0.6em] Treated $\times$ 2022--23 & 0.031*** & 0.031** & -0.001 & 37.7** & 60.3*** & 43.1* \\ \addlinespace[0.3em] & (0.010) & (0.012) & (0.024) & (17.8) & (15.6) & (22.2) \\ \addlinespace[0.6em] Mean DV & 0.188 & 0.108 & 0.399 & 84.6 & 67.7 & 93.9 \\ \addlinespace[0.3em] R-squared & 0.107 & 0.080 & 0.114 & 0.211 & 0.247 & 0.252 \\ \addlinespace[0.3em] Observations & 56,003 & 38,546 & 16,258 & 10,644 & 3,866 & 6,192 \\ \hline  \end{tabular}

\begin{tablenotes}
\footnotesize
\item *** $p<0.01$, ** $p<0.05$, * $p<0.01$. This table presents the difference-in-differences estimates based on Specification \eqref{eq:basicDiD}. Moved House is an indicator variable taking value of 1 if respondents moved residential address in the survey year, 0 otherwise; individuals who relocate across states are excluded. Distance to CBD is calculated for movers only, based on the postcode-level distance between CBD of the state capital city and movers' new address. All regressions include state-by-occupation fixed effects and survey wave fixed effects. Standard errors are clustered at the state-by-occupation level. The mean of the dependent variable is based on the average value in control states in 2020.
\end{tablenotes}
\end{threeparttable}
\end{table}

\begin{landscape}
\begin{table}[htbp]
\small
\begin{threeparttable}
\caption{Employees' response: COVID anxiety and health conditions } \label{tab: anxiety}
\begin{tabular}{lcccccc}              \hline \hline & (1) & (2) & (3) & (4) & (5) & (6) \\ \hline \addlinespace \multicolumn{7}{c}{Panel A: COVID-Related Variables (2022--23 only)}             \\ \addlinespace & Anxiety & Anxiety & Crowd Stress & Crowd Stress & Infected Last 12m & Infected Ever \\ & (1-7) & ($\ge 4$) & (1-7) & ($\ge 4$) & (0/1) & (0/1) \\ \cmidrule(lr){2-7} Treated & 0.235*** & 0.057*** & 0.125*** & 0.027*** & 0.001 & 0.052*** \\ \addlinespace[0.3em] & (0.049) & (0.012) & (0.039) & (0.011) & (0.006) & (0.019) \\ \addlinespace[0.6em] Observations & 4,991 & 4,991 & 4,990 & 4,990 & 1,655 & 2,549 \\ \addlinespace[0.3em] \hline \addlinespace \multicolumn{7}{c}{Panel B: Other Health-Related Variables}             \\ \addlinespace & General Health & Physical Difficulty & Mental Difficulty & Long-term Cond. & Short Breath & Hours worked/Week \\ & (1-5) & (0/1) & (0/1) & (0/1) & (0/1) & (cont.) \\ \cmidrule(lr){2-7} Treated $\times$ 2020--21 & -0.028 & -0.008 & 0.016* & -0.009 & -0.034 & 0.283 \\ \addlinespace[0.3em] & (0.029) & (0.008) & (0.009) & (0.013) & (0.023) & (0.364) \\ \addlinespace[0.6em] Treated $\times$ 2022--23 & 0.006 & 0.008 & 0.003 & -0.028** & 0.007 & -0.359 \\ \addlinespace[0.3em] & (0.036) & (0.006) & (0.012) & (0.013) & (0.022) & (0.315) \\ \addlinespace[0.6em] Observations & 52,220 & 52,408 & 52,397 & 57,883 & 7,342 & 57,896 \\ \hline \addlinespace \multicolumn{7}{c}{Panel C: Related Household Expenditures}             \\ \addlinespace & Health Providers & Medicines & Cigarette & Meals Out & Pub Transport/WH & Fuel/WH \\ & (\textdollar) & (\textdollar) & (\textdollar) & (\textdollar) & (\textdollar) & (\%) \\ \cmidrule(lr){2-7} Treated $\times$ 2020--21 & -88.459*** & -53.962*** & 246.357*** & -393.480*** & -0.232*** & -0.276*** \\ \addlinespace[0.3em] & (31.406) & (19.661) & (55.460) & (99.939) & (0.034) & (0.060) \\ \addlinespace[0.6em] Treated $\times$ 2022--23 & -113.596** & -70.455*** & 197.540*** & -150.834 & -0.038 & -0.089 \\ \addlinespace[0.3em] & (47.909) & (22.924) & (51.517) & (119.578) & (0.031) & (0.062) \\ \addlinespace[0.6em] Observations & 49,395 & 49,395 & 51,611 & 51,611 & 51,611 & 51,611 \\ \hline \end{tabular}

\begin{tablenotes}
\footnotesize
\item *** $p<0.01$, ** $p<0.05$, * $p<0.01$. Panel A of this table uses variables only surveyed after the pandemic. Anxiety and crowd stress (Columns 1 and 3) are rated from 1 (none) to 7 (extreme), collected in 2022–23; Columns 2 and 4 use binary indicators for levels above 4. COVID infection outcomes (Columns 5 and 6) are from the 2022 wave. All regressions control for wave fixed effects, individual, occupation, and industry characteristics. State-by-occupation fixed effects are excluded to capture overall treatment effects; standard errors are clustered at the state-by-occupation level. Panels B and C use the Difference-in-Differences estimation as specified in \eqref{eq:basicDiD}.  Panel B includes self-rated health (1 = excellent, 5 = poor) in Column 1; Columns 2–5 use indicators for physical, mental, or long-term health issues, and shortness of breath (conditional on having a long-term condition). Column 6 measures weekly working hours. Panel C reports household expenditures, winsorized at the 99th percentile; transport and fuel spending are adjusted for working hours.
\end{tablenotes}
\end{threeparttable}
\end{table}
\end{landscape}

{
\clearpage
\pagestyle{fancy}
\begin{appendices}
\setcounter{figure}{0}
\renewcommand{\thefigure}{A\arabic{figure}}
\setcounter{table}{0}
\renewcommand{\thetable}{A\arabic{table}}
\setcounter{equation}{0}
\renewcommand{\theequation}{A\arabic{equation}}
\setcounter{footnote}{0}
\renewcommand{\thefootnote}{A\arabic{footnote}}
\setcounter{page}{1}
\renewcommand{\thepage}{A\arabic{page}}
\setcounter{tocdepth}{1}
\addtocontents{toc}{\protect\setcounter{tocdepth}{1}}
\begin{center}
{\Large \textbf{Web Appendix for ``{\papertitle{}}''}} \\ \medskip

\author{Laura Ketter\space\space\space\space\space\space \space\space\space\space\space\space 
Todd Morris
\space\space\space\space\space\space \space\space\space\space\space\space 
Lizi Yu
\vspace{-0.1in}
} 
 \end{center}

\captionsetup[section]{list=yes}
\captionsetup[figure]{list=yes}
\captionsetup[table]{list=yes}
{
\thispagestyle{empty}
\footnotesize
\setstretch{1}
 \tableofcontents
 \listoffigures
 \listoftables
}

\newpage
\begin{table}[htbp]
\small \centering
\caption{Summary statistics~--- Baseline sample} \label{tab: summarystats}
\begin{threeparttable}
\begin{tabular}{l R{1.5cm}} 
\hline\hline
                    &        Mean\\
\hline \addlinespace
Age                 &       38.3\\
\quad (Std. dev.) & (10.9)\\ \addlinespace[0.2em]
Female              &        51.1\% \\
Partnered           &        71.0\%\\
University degree   &        45.7\%\\
Hours worked per week&       39.0\\ 
\quad (Std. dev.) & (11.1)\\ 
\addlinespace
\underline{Top three occupations}:&           \\
$\quad$ Professionals&        33.7\%\\
$\quad$ Clerical and Administrative Workers&        30.8\%\\
$\quad$ Managers    &        24.7\%\\ \addlinespace
\underline{WFH Measures}:       &           \\
$\quad$ WFH hours   &        4.5 \\
\qquad (Std. dev.) & (10.0)\\ \addlinespace[0.2em]
$\quad$ WFH share   &        11.9\% \\
$\quad$ Any WFH     &        32.1\% \\
$\quad$ WFH with agreement&        16.4\%\\
$\quad$ WFH share $\ge$ 0.6&        8.2\% \\
$\quad$ WFH share = 1&        5.2\%\\ \addlinespace
\underline{State of residence}: &           \\
$\quad$ New South Wales&        29.6\%\\
$\quad$ Victoria    &        27.8\%\\
$\quad$ Queensland  &        19.1\%\\
$\quad$ Western Australia&        9.2\%\\
$\quad$ South Australia&        7.4\%\\
$\quad$ Australian Capital Territory&        4.7\%\\
$\quad$ Tasmania    &        1.4\%\\
$\quad$ Northern Territory&        0.8\%\\ \addlinespace
Observations & 57,896 \\
\hline
\end{tabular}

\begin{tablenotes}
\footnotesize
\item This table presents summary statistics for the main analysis sample (office workers in major urban areas from the HILDA survey). The six variables capturing WFH practices are: (i) share of total hours WFH; (ii) total WFH hours in respondent's main job; (iii) an indicator for any WFH; (iv) an indicator for WFH under a formal agreement; (v) an indicator for working from home $\geq$60\% of total hours; and (vi) an indicator for fully remote work.
\end{tablenotes}
\end{threeparttable}
\end{table}
\clearpage
\begin{table}[htbp]
\small
\begin{threeparttable}
\caption{Dose-response relationship~--- Differences-in-differences estimates by state} \label{tab: dose}
\begin{tabular}{lcccccc}              \hline \hline & WFH & WFH & Any & WFH with & WFH & WFH \\ & share & hours & WFH & agreement & share $\geq 0.6$ & share $=1$ \\ & (1) & (2) & (3) & (4) & (5) & (6) \\ \hline \addlinespace 
\multicolumn{7}{l}{\underline{Treatment effects in 2020--21}:} \\		\addlinespace[0.6em]									
\quad ACT	&	0.112***	&	4.7***	&	0.106***	&	0.099***	&	0.126***	&	0.087***	\\	\addlinespace[0.3em]
	&	(0.028)	&	(1.0)	&	(0.022)	&	(0.031)	&	(0.035)	&	(0.023)	\\	\addlinespace[0.6em]
\quad NSW	&	0.241***	&	9.6***	&	0.157***	&	0.198***	&	0.273***	&	0.246***	\\	\addlinespace[0.3em]
	&	(0.022)	&	(0.9)	&	(0.022)	&	(0.024)	&	(0.023)	&	(0.020)	\\	\addlinespace[0.6em]
\quad VIC	&	0.349***	&	13.6***	&	0.252***	&	0.248***	&	0.375***	&	0.372***	\\	\addlinespace[0.3em]
	&	(0.030)	&	(1.1)	&	(0.027)	&	(0.024)	&	(0.033)	&	(0.034)	\\	\addlinespace[0.6em]
\qquad \(p\)-value: ACT = NSW	&	\(<\)0.001	&	\(<\)0.001	&	0.020	&	0.002	&	\(<\)0.001	&	\(<\)0.001	\\	\addlinespace[0.8em]
\qquad \(p\)-value: NSW = VIC	&	\(<\)0.001	&	\(<\)0.001	&	\(<\)0.001	&	0.001	&	\(<\)0.001	&	\(<\)0.001	\\	\addlinespace[0.8em]

\multicolumn{7}{l}{\underline{Treatment effects in 2022--23}:} \\ \addlinespace[0.6em]							
\quad ACT	&	0.021	&	0.1	&	0.035	&	0.057**	&	0.014	&	0.011	\\	\addlinespace[0.3em]
	&	(0.020)	&	(1.0)	&	(0.024)	&	(0.026)	&	(0.026)	&	(0.016)	\\	\addlinespace[0.6em]
\quad NSW	&	0.089***	&	3.1***	&	0.072***	&	0.082***	&	0.119***	&	0.022**	\\	\addlinespace[0.3em]
	&	(0.015)	&	(0.6)	&	(0.021)	&	(0.017)	&	(0.020)	&	(0.011)	\\	\addlinespace[0.6em]
\quad VIC	&	0.123***	&	4.5***	&	0.128***	&	0.129***	&	0.134***	&	0.044***	\\	\addlinespace[0.3em]
	&	(0.016)	&	(0.6)	&	(0.021)	&	(0.016)	&	(0.020)	&	(0.012)	\\	\addlinespace[0.6em]
\qquad \(p\)-value: ACT = NSW	&	\(<\)0.001	&	0.002	&	0.182	&	0.344	&	\(<\)0.001	&	0.358	\\	\addlinespace[0.8em]
\qquad \(p\)-value: NSW = VIC	&	\(<\)0.001	&	\(<\)0.001	&	0.016	&	\(<\)0.001	&	0.082	&	0.096	\\	\addlinespace[0.8em]
R-squared	&	0.434	&	0.452	&	0.264	&	0.284	&	0.370	&	0.329	\\	\addlinespace[0.8em]
Observations	&	57,896	&	57,896	&	57,896	&	57,896	&	57,896	&	57,896	\\	



\hline 
\end{tabular}

\begin{tablenotes}
\footnotesize
\item *** $p<0.01$, ** $p<0.05$, * $p<0.01$. This table presents the difference-in-differences estimates from a modified version of Specification~\eqref{eq:basicDiD} that allows for different short- and medium-term effects in each of the three treated states. All regressions include state-by-occupation fixed effects, survey wave fixed effects and rich period-specific controls for industry, occupation and other factors affecting WFH. Standard errors in parentheses are clustered at the state-by-occupation level.
\end{tablenotes}
\end{threeparttable}
\end{table}


\clearpage
\begin{figure}[htbp]
\centering
\caption{The rise of WFH in Australia by occupation} \label{fig:aus_wfh_share_office_nonoffice}
\includegraphics[width=\textwidth]{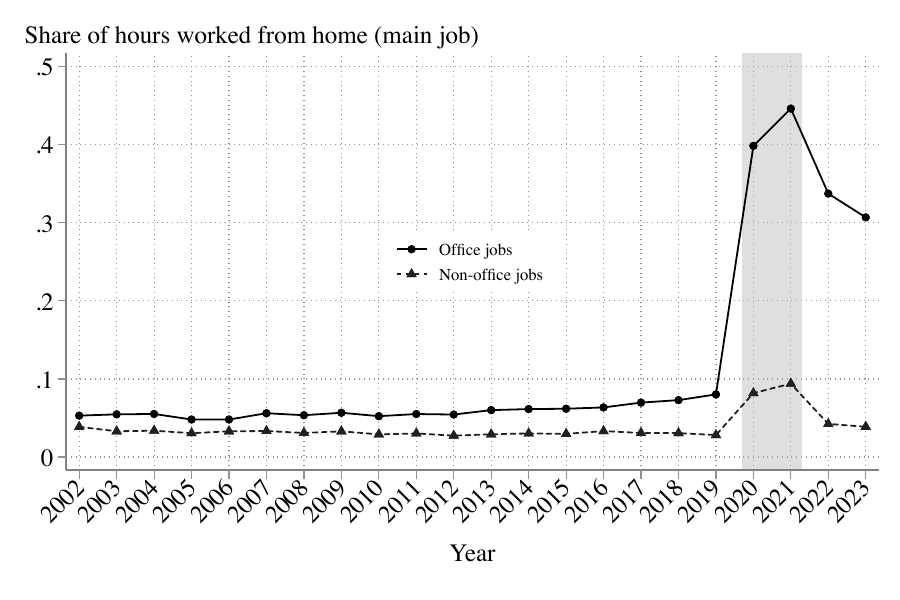}
\begin{minipage}{\textwidth}
{\footnotesize \underline{Notes}: This figure plots the share of hours worked from home from 2002 to 2023 from the HILDA survey, separately for office and non-office workers for all employees in major urban cities. The WFH share is measured as the proportion of hours worked from home relative to total hours worked in the main job. Office workers are primarily composed of clerical and administrative workers, managers, and professionals.} 
\end{minipage}
\end{figure}

\clearpage
\begin{figure}[htbp]
    \centering
    \caption{Trends in office vacancy rates in  central business districts of each state}
    \label{fig:office_vacancy_rates_det}
   \includegraphics[width=\textwidth]{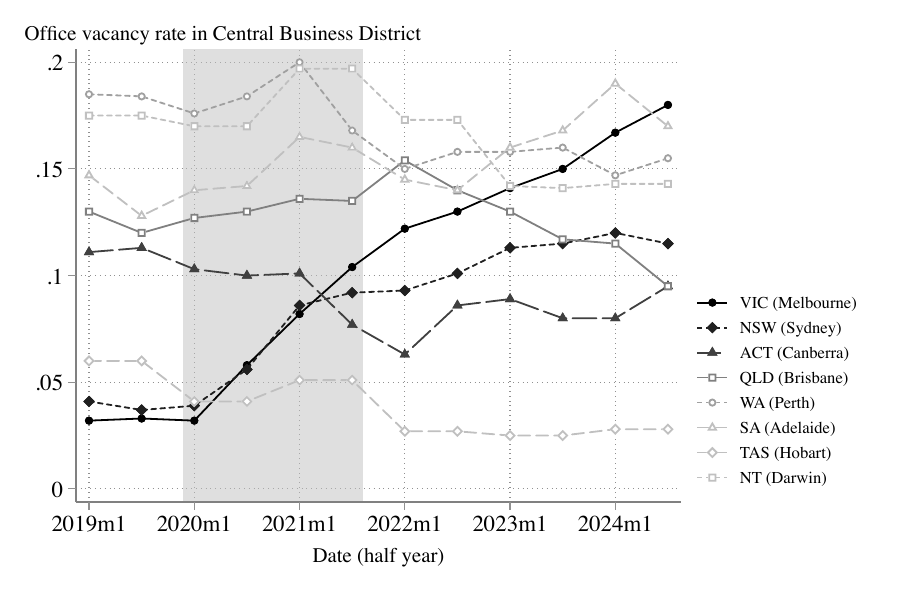}

    \begin{minipage}{\textwidth}
{\footnotesize \underline{Notes}: This figure plots the office vacancy rate in the capital cities of each state and territory from 2019 to 2024, using data from the \href{https://www.propertycouncil.com.au/}{Property Council of Australia}. The measure captures the proportion of unoccupied office space in central business districts. The gray shaded area shows the period where lockdowns were in place across parts or all of Australia.} 
\end{minipage}
\end{figure}

\clearpage
\begin{figure}[htbp]
    \centering
    \caption{Examining adoption of home-office technologies/infrastructure via Google Trends data~--- Relative search interest for different terms in all states}
    \label{fig:google_trends_app2}
    \begin{subfigure}[t]{0.49\textwidth} \caption{``Computer''}\includegraphics[width=\textwidth]{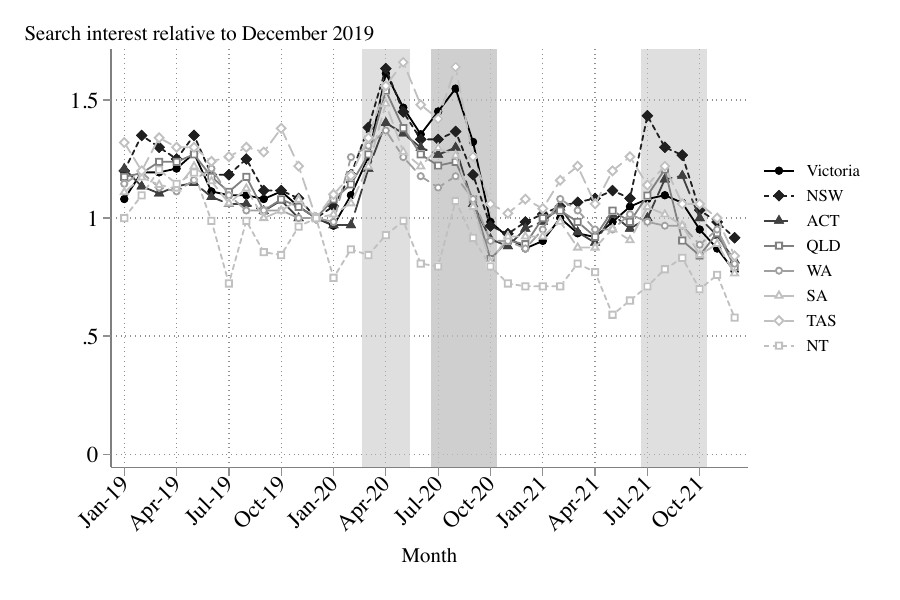}   
    \end{subfigure}
    \begin{subfigure}[t]{0.49\textwidth}
    \caption{``Computer monitor''}\includegraphics[width=\textwidth]{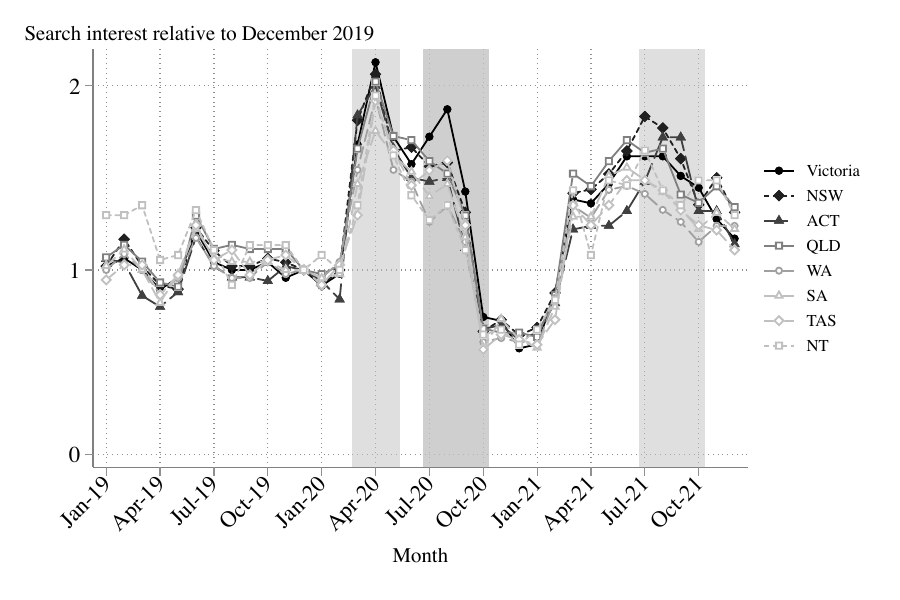}   
    \end{subfigure} \vspace{0.1cm}

       \begin{subfigure}[t]{0.49\textwidth} \caption{``Desk''}\includegraphics[width=\textwidth]{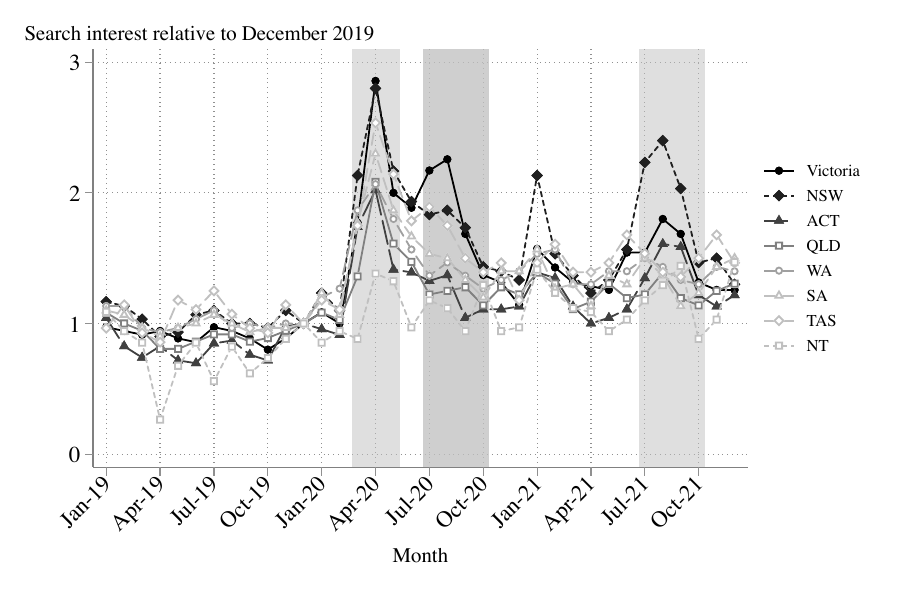}   
    \end{subfigure}
    \begin{subfigure}[t]{0.49\textwidth}
    \caption{``Office chair''}\includegraphics[width=\textwidth]{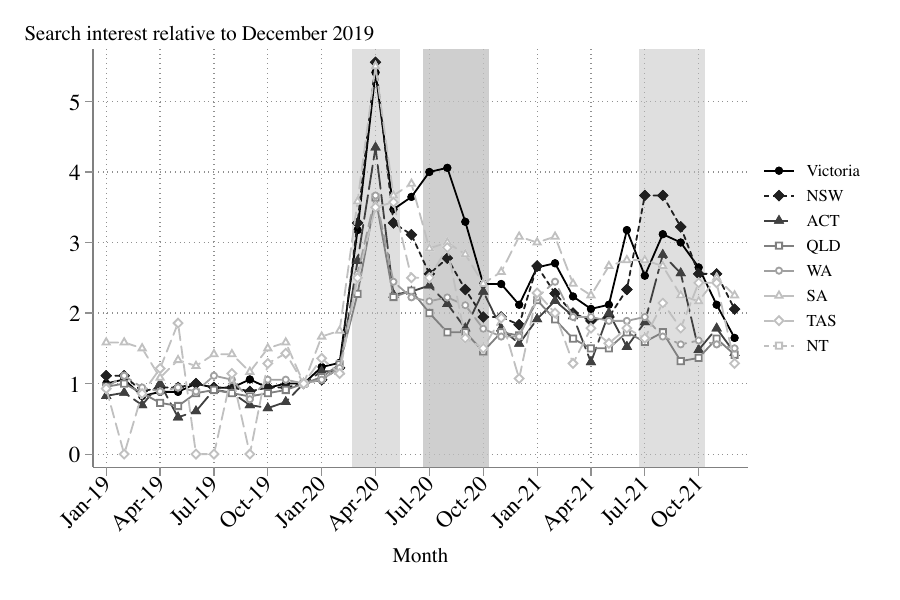}   
    \end{subfigure} \vspace{0.1cm}

   \begin{subfigure}[t]{0.49\textwidth} \caption{``Zoom''}\includegraphics[width=\textwidth]{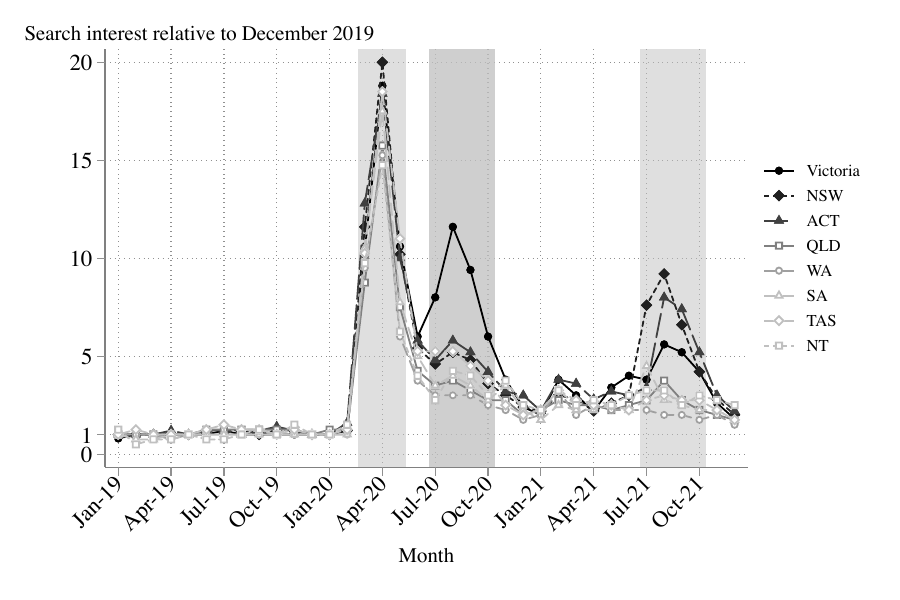}   
    \end{subfigure}
    \begin{subfigure}[t]{0.49\textwidth}
    \caption{``Microsoft Teams''}\includegraphics[width=\textwidth]{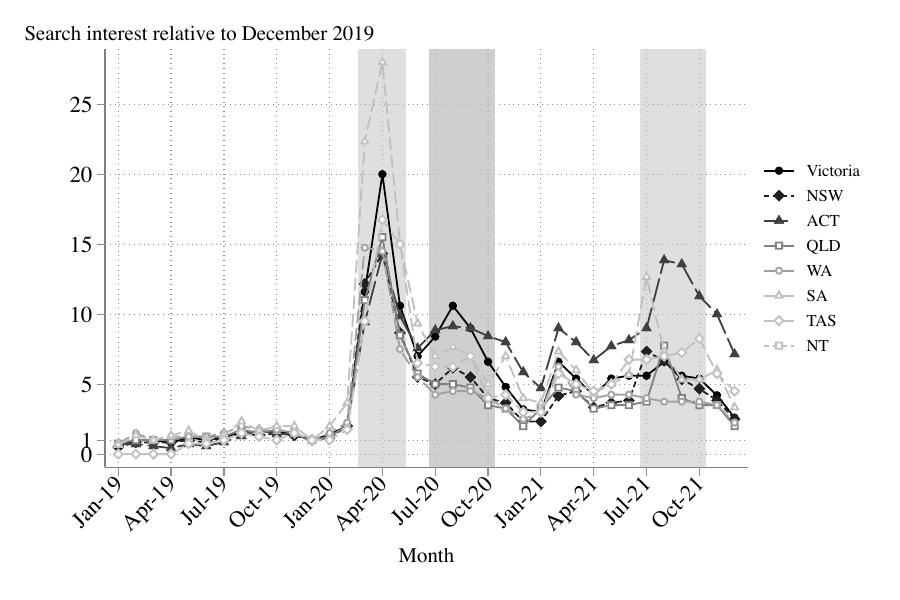}   
    \end{subfigure} 

 \begin{minipage}{\textwidth}
{\footnotesize \underline{Notes}: These figures plot search interest for each specified keyword from 2019 to 2021, using data from \hyperref[https://trends.google.com/trends/]{Google Trends}. We weight the state-level monthly indices from Google Trends in the control states by the population size of the largest city in each state. We then normalize our indices by the relative amount of search interest in the region in December 2019.} 
\end{minipage}
\end{figure}

\clearpage
\section{More Robustness: Alternative Estimations}\label{more robustness}
Next, we conduct a variety of robustness checks to affirm the stability and validity of our results, as presented in Figure \ref{fig: robust}. Below we summarize several key robustness exercises and their outcomes.

\noindent\textbf{Alternative Fixed Effects and Specification Choices.} Our baseline DiD specification includes a set of state-by-occupation fixed effects to account for any persistent differences in WFH propensity across occupations and states. We test alternative FE structures to ensure our results are not sensitive to this choice, and standard errors are clustered at the corresponding fixed-effect level in each case. Our results are robust to replacing state-by-occupation fixed effects with state-by-industry FEs, or worker individual fixed effects, or finer geographic FEs at the postcode level. In all these robustness exercises, the estimated treatment effects remain robust and significant, with only minimal changes in coefficients.

\noindent\textbf{Inference and Clustering.} In our main specification, we cluster standard errors at the state-by-occupation level to account for potential correlations in WFH behaviors among workers within the same state and occupation group. We further assess the robustness of our inference by clustering standard errors at the broader state/territory level (8 clusters), and by applying a wild cluster-bootstrap procedure \citep{cameron2008bootstrap} to address concerns of the small number of clusters.\footnote{We include confidence intervals with and without jackknifing of the data generation process, as suggested by \cite{mackinnon2023cluster}.}  Reassuringly, even with these very conservative exercises, our key estimates remain highly statistically significant~--- e.g., the lockdown-period and post-period treatment effects are statistically significant at the 1\% level for WFH share.

\noindent\textbf{Alternative Treatment Definitions.} Our main analysis defines ``treated" workers based on the state in which they are observed at each survey interview (i.e., their current location). One concern might be that some workers moved between states during/after the pandemic, potentially diluting the treatment and control distinction. To address this, we re-define treatment status based on the individual's state of residence in 2019 and then tracked outcomes regardless of subsequent moves.\footnote{This results in the loss of 17\% of observations, since individuals who leave the survey prior to 2019 or join after 2019 are not included.} Again, the DiD results remain very similar under this alternative definition, because interstate migration in our sample was quite low~--- only 2.7\% of workers moved between treated and control states from 2019 to 2022.

\noindent\textbf{Alternative Samples.} In the first two practices, we restrict the sample to address the concerns discussed in Section \ref{robustness}, by limiting the analysis to postcodes with comparable pre-COVID commute times only, and to capital cities only. In both cases, the estimated treatment effects remain highly similar. In our main analysis, we focus on office workers living in major urban areas in Australia. This helps improve comparability between treated and control regions and ensures that the analysis is concentrated on workers for whom WFH is most relevant. As a robustness check, we relax this restriction and re-estimate the regressions using two broader samples: all urban employees living in major cities (including those with non-office jobs), and all urban employees (including those in small urban areas). As expected, the estimated coefficients become attenuated when expanding the sample to include jobs less suitable for WFH, and smaller regional areas where remote work is less feasible or relevant. Nevertheless, the estimated effects still remain statistically significant in these alternative samples.

\newpage
    \begin{figure}[!htbp]
\centering
\caption{Robustness of DiD estimates to alternative specifications and samples} \label{fig: robust}
\begin{subfigure}[t]{0.49\textwidth}
\caption{WFH share}
\includegraphics[width=\textwidth]{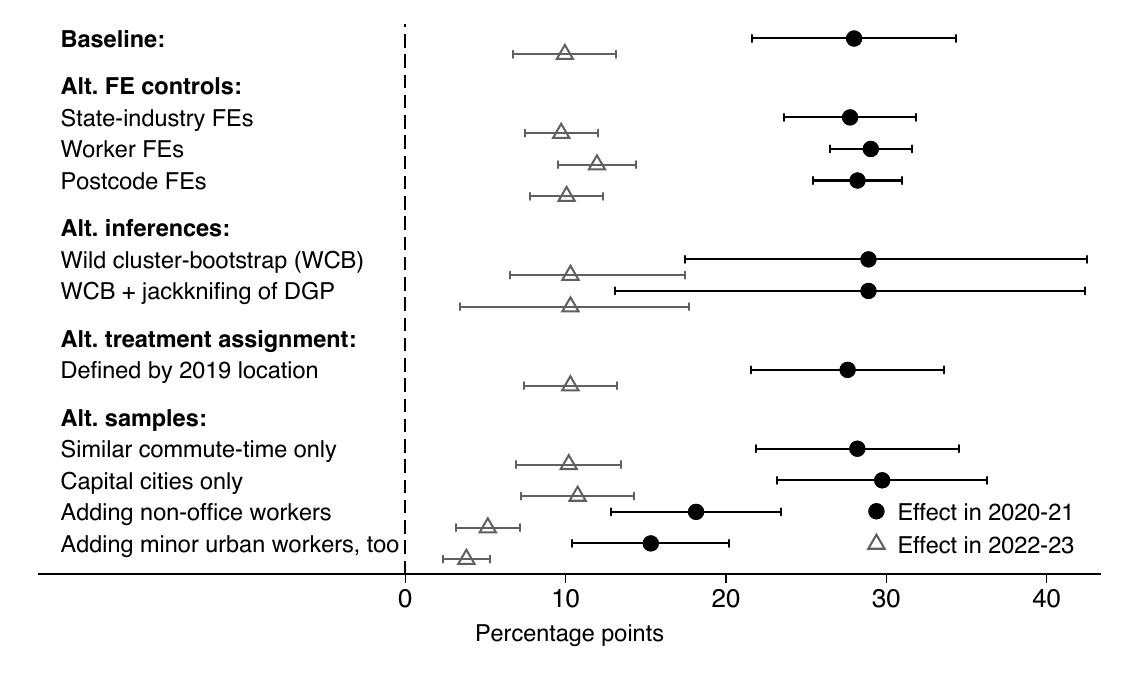} 
\end{subfigure}
\begin{subfigure}[t]{0.49\textwidth}
\caption{WFH hours}
\includegraphics[width=\textwidth]{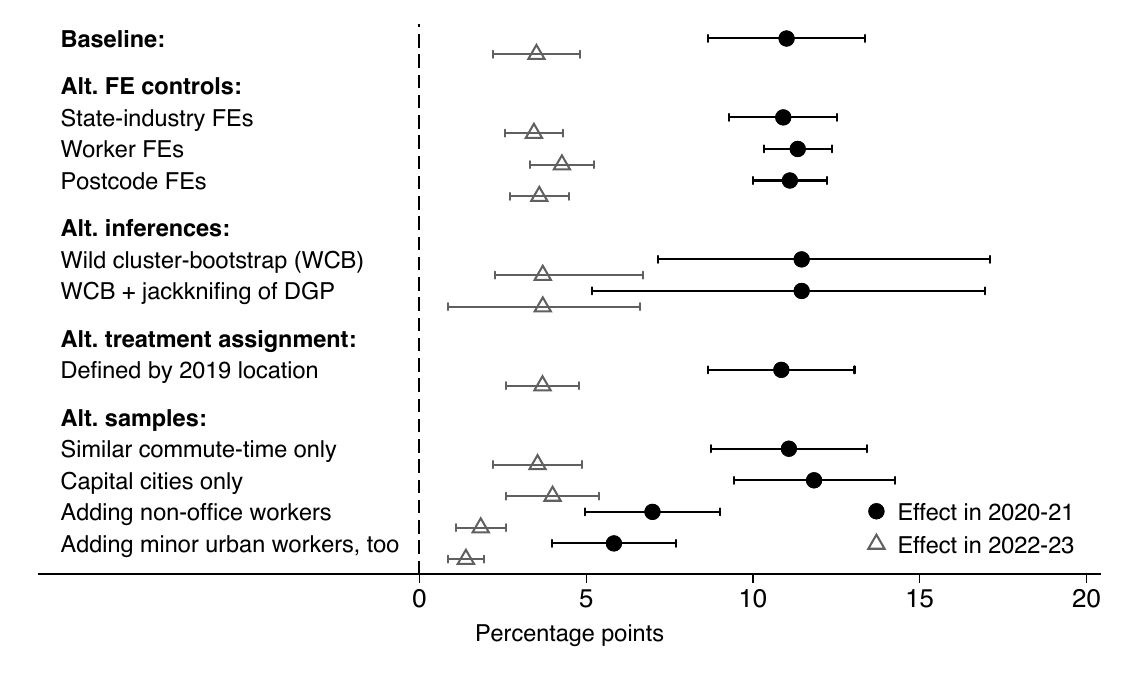} 
\end{subfigure}
\vspace{1cm}

\begin{subfigure}[t]{0.49\textwidth}
\caption{Any WFH}
\includegraphics[width=\textwidth]{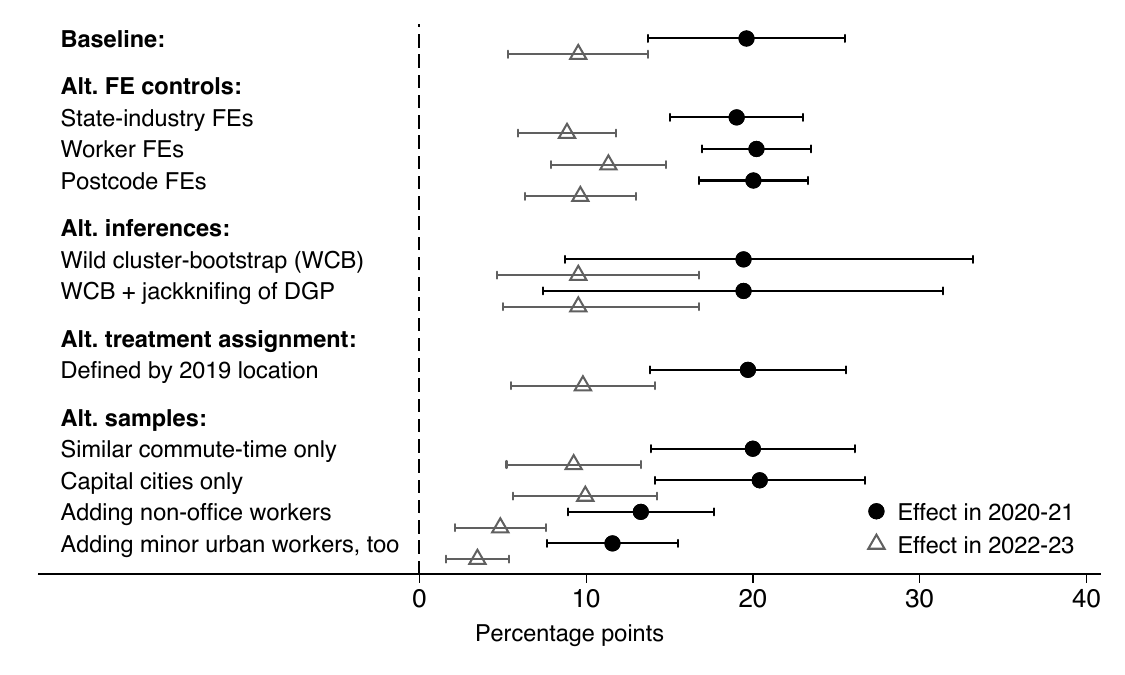} 
\end{subfigure}
\begin{subfigure}[t]{0.49\textwidth}
\caption{Any WFH with formal agreement}
\includegraphics[width=\textwidth]{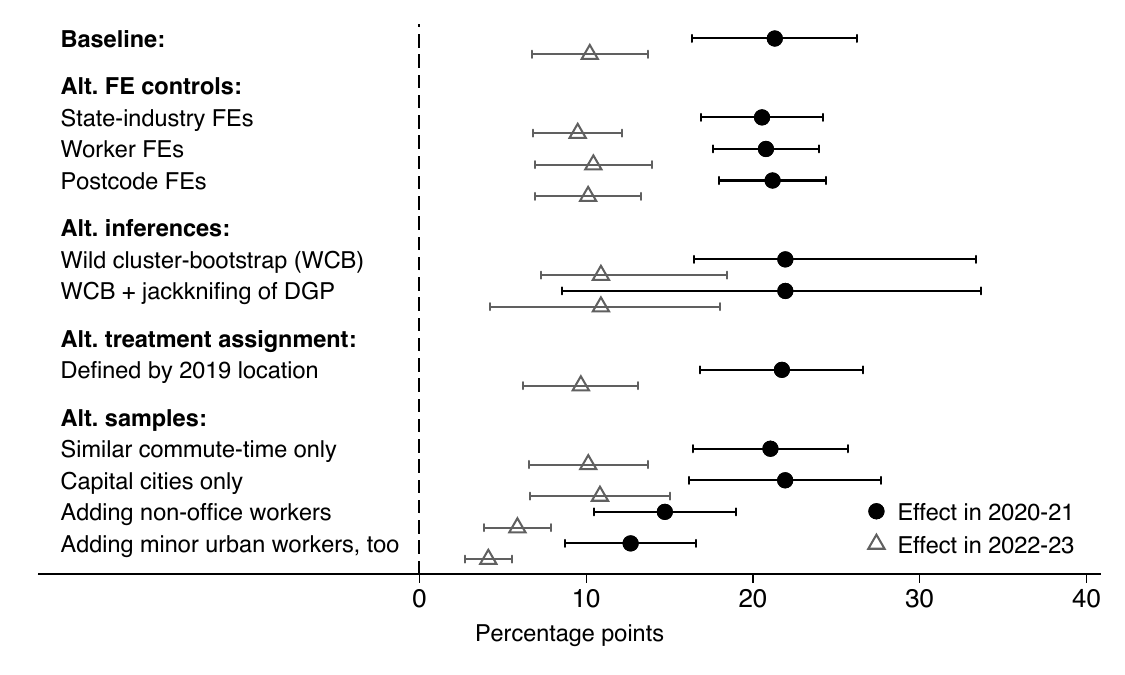} 
\end{subfigure} 
\vspace{1cm}

\begin{subfigure}[t]{0.49\textwidth}
\caption{Mostly WFH (WFH share $\geq$ 0.6)}
\includegraphics[width=\textwidth]{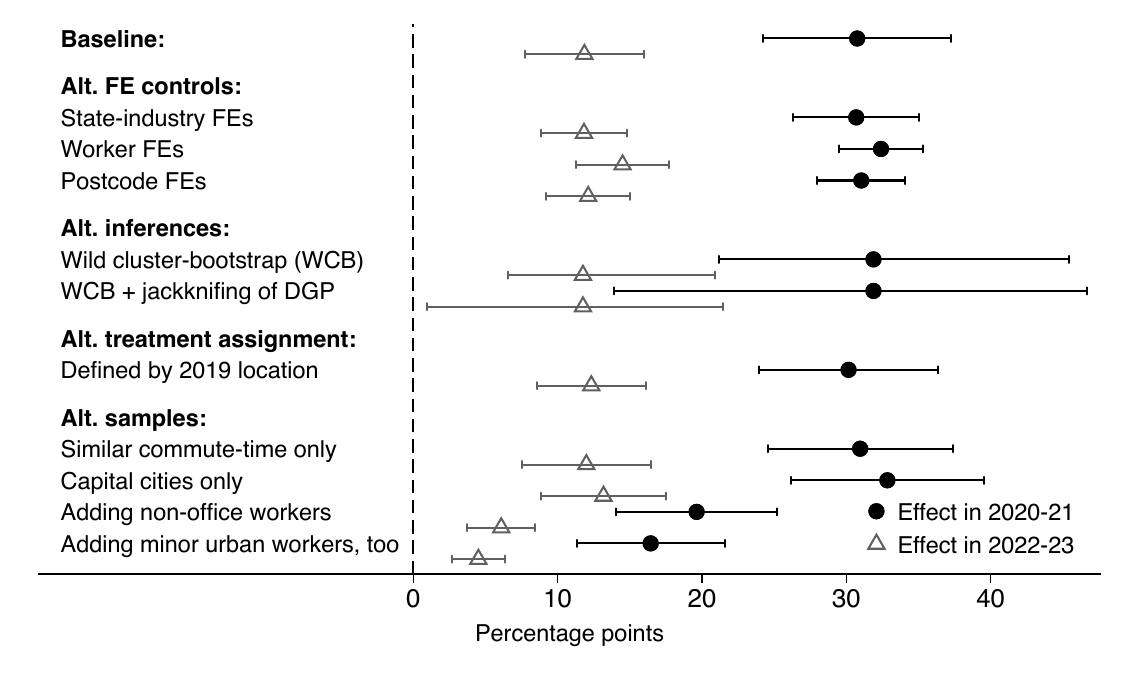} 
\end{subfigure}
\begin{subfigure}[t]{0.49\textwidth}
\caption{Fully remote (WFH share $=$ 1)}
\includegraphics[width=\textwidth]{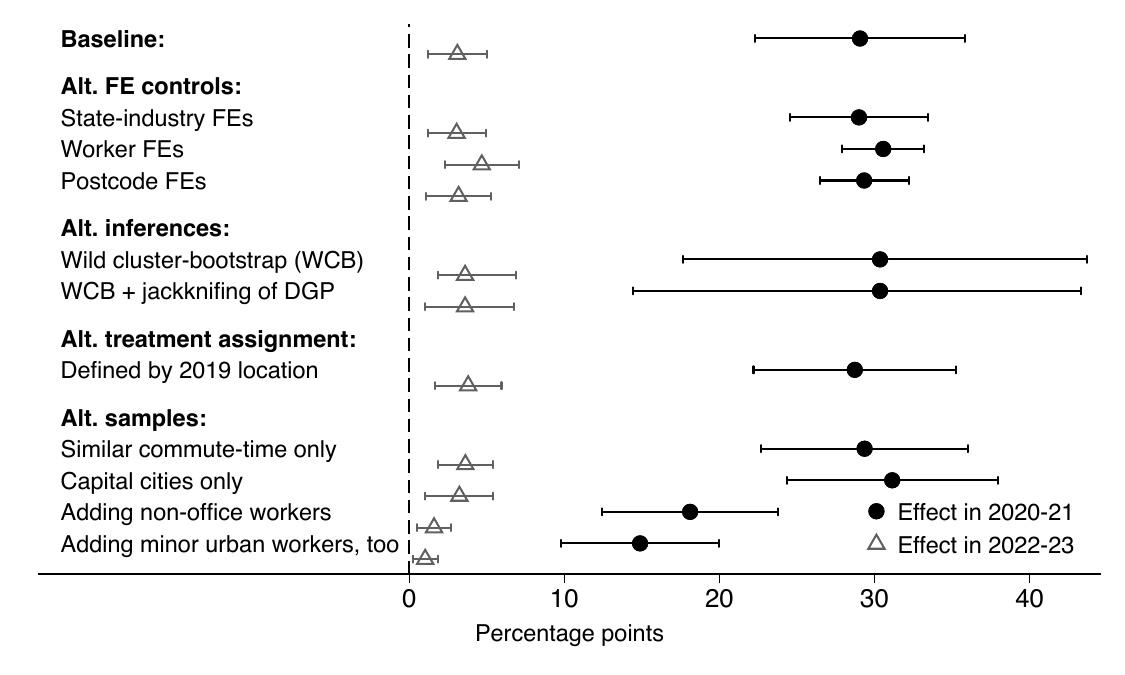} 
\end{subfigure}

\begin{minipage}{\textwidth}
{\footnotesize \underline{Notes}: These figures demonstrate the robustness of the estimated effects to alternative empirical specifications and samples. We present point estimates and 95\% confidence intervals of Specification~\eqref{eq:basicDiD} with different modeling choices. Specifically, we replace the state-by-occupation fixed effects with state-by-industry fixed effects, worker fixed effects, and post code fixed effects, respectively. We also cluster standard errors at the state/territory level and apply wild cluster-bootstrap methods \citep{cameron2008bootstrap} to account for the small number of clusters. Additionally, we define treatment and control status based on workers’ 2019 locations and present estimates across several sample definitions: restricting the sample of workers in control regions to those in postcodes with similar average commuting times as workers in treated regions, restricting to workers in capital cities, including non-office workers, and including all urban employees nationwide.} 
\end{minipage}
\end{figure}
\clearpage
\section{Heterogeneity by employee characteristics} \label{more heterogeneity}
In Figure \ref{fig: correlation}, we plot the correlation between the immediate WFH effect during the 2020-21 lockdown and the medium-run WFH effect observed in 2022-23, across demographic groups. It reveals a clear positive correlation: subgroups that experience larger lockdown-induced increases in WFH tend to retain higher WFH rates post-pandemic. Such a positive correlation in the estimates across subgroups may indicate the presence of habit formation among employees and employers \citep{charness2009incentives, fujiwara2016habit, ito2018moral, moore2024shaping}, although the other mechanisms uncovered (employer responses, technological investments, and relocation) may also contribute to this positive relationship. Graphically, points cluster along an upward-sloping trajectory~--- those far to the right (high initial WFH uptake) are generally higher up (high post-COVID WFH). The outcomes that exhibit higher levels of correlation are any WFH, WFH share, WFH $\ge$ 0.6, and WFH hours in main job, suggesting that the persistence of WFH is most pronounced among subgroups that adopt at least some level of remote work during lockdown. 

\noindent \textbf{University-Educated vs. Less Educated Workers.} One striking trend is the strong and sustained WFH uptake among university-educated workers. These workers have among the highest jumps in WFH during lockdown and also exhibit some of the largest medium-run WFH levels. In the scatter, the university-educated subgroup sits toward the top-right, signaling both high immediate adoption and strong persistence. In contrast, less-educated office workers see more modest WFH gains and correspondingly smaller persistence, keeping their points nearer the origin. This is unsurprising since university graduates are concentrated in occupations more suitable for remote work, which allow them to transit effectively during lockdown and retain a substantial WFH advantage post COVID. These sizable gaps across all major WFH metrics indicate that educated white-collar professionals not only lead the way in initial WFH adoption but also normalized hybrid work in the new equilibrium. 

\noindent  \textbf{Gender Differences (Women vs. Men).} Gender differences in Figure \ref{fig: correlation} exhibit a notable divergence in WFH persistence. During the lockdown, men and women increased WFH at similar rates (their points have comparable $x$-axis values), reflecting the broad, mandatory nature of WFH at that time. But in the post-COVID period, the female subgroup clearly stands above the male subgroup on the $y$-axis, indicating that women sustain WFH at higher levels. Women in treated states are about 10 pp more likely to have any WFH in 2022--23 relative to their male counterparts; and they also maintain greater intensive use of WFH (e.g. higher WFH share, WFH hours and more frequent 3+ WFH days) than men, suggesting that remote work became an established part of women's work habits. This gender gap in persistence aligns with broader findings that women especially value flexible work arrangements and the ability to better balance work and home responsibilities. 

\noindent  \textbf{Parents of Young Children, Partnered and Single Workers.} A perhaps counterintuitive finding in Figure \ref{fig: correlation} is the lack of a large differential effect for parents with young children (aged 0--9). One might expect parents of toddlers or school-aged children to show an especially high WFH uptake (during school closures and daycare disruptions) and a strong continued use of WFH afterward for caregiving convenience. However, the subgroup of workers with young dependents does not stand out with an unusually high point.  While parents of young children experience a substantial initial increase in WFH, as it is for nearly all office workers, their uptake is not significantly higher than that of non-parents. More notably, their WFH persistence into the medium run is actually lower than that of other partnered workers without young children or even single workers. This suggests that while caregiving constraints do not translate into long-term remote work adoption at higher rates than other demographic groups. This result likely reflects two factors. First, during lockdown the shift to home was nearly universal among white-collar employees, so parents and non-parents alike were forced into WFH regardless of family status. Second, by the post-COVID period, schools and childcare centers had reopened, alleviating much of the pressure that would uniquely compel parents to remain home. 

\noindent  \textbf{Age Group Dynamics.} Age-based patterns also emerge from the subgroup comparisons. show both strong WFH uptake during the lockdown and notable persistence post-pandemic. In each panel, this group typically appears in the upper-right quadrant, indicating a substantial initial shift to remote work and a sizable portion retained afterward. The slightly younger cohort, aged 35--44, also demonstrates a strong immediate WFH response, but with more variation in medium-run persistence across WFH measures. Workers in their 30s and early 40s are often in their prime career stage~--- many hold senior professional or managerial positions that are amenable to remote work, and they have the experience and authority to continue WFH effectively.

In contrast, points of younger employees (18--24 and 25--34) fall toward the lower-right of each panel. While they have worked from home extensively in 2020--21, they appear to have lower medium-run WFH levels. Younger workers may have been more inclined to return to office life for mentorship, networking. They also often occupy junior roles that might require in-person training or collaboration, limiting sustained WFH.  Meanwhile, older workers (55--59) consistently appear in the lower-left corner across all outcomes, suggesting both limited WFH uptake during the lockdown and minimal persistence afterward, possibly due to more ingrained work habits.  

Overall, the scatterplot matrix in Figure \ref{fig: correlation} exhibits a broadly positive relationship between the immediate and medium-run effects of the pandemic on remote work across diverse groups. WFH persistence appears particularly strong among women, university-educated workers, and mid-career cohorts.

\clearpage
\newpage
\begin{figure}[htbp]
\centering
\caption{Heterogeneity in the short- and medium-term effects on WFH share by occupation} \label{fig: heterogeneity}
\begin{subfigure}[t]{0.49\textwidth}
\caption{Short- and medium-term treatment effects}
\includegraphics[width=\textwidth]{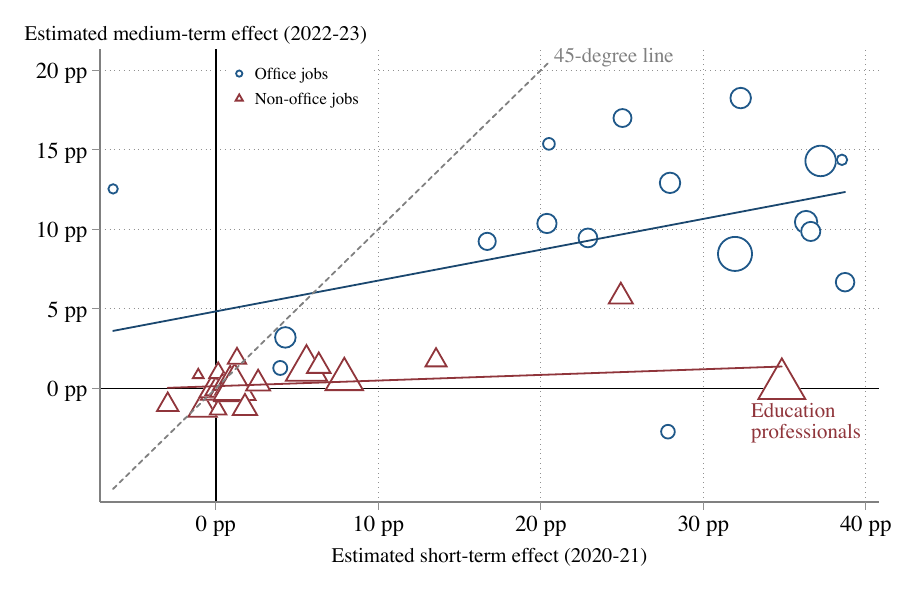}
\end{subfigure}

\begin{subfigure}[t]{0.49\textwidth}
\caption{Short-term effects and pre-COVID levels}
\includegraphics[width=\textwidth]{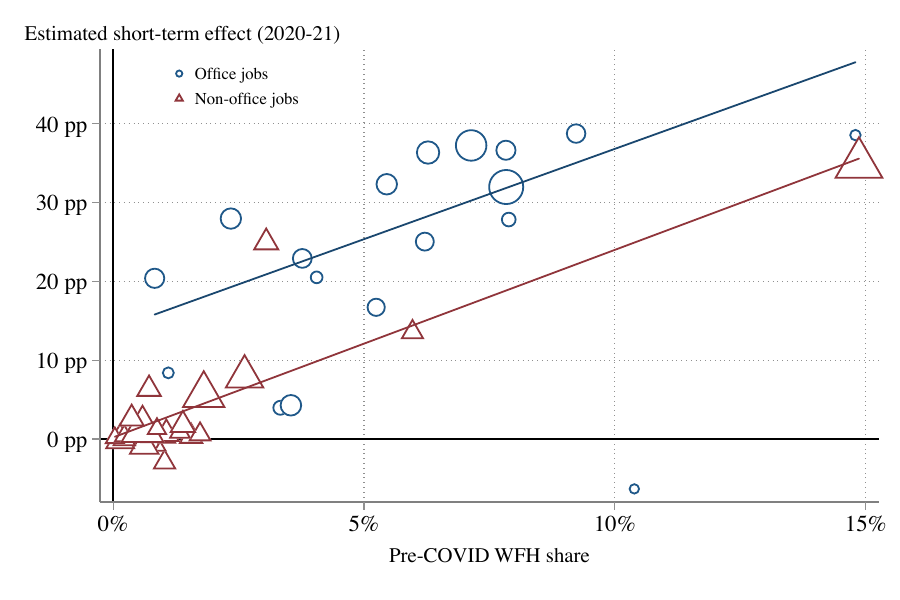} 
\end{subfigure}
\begin{subfigure}[t]{0.49\textwidth}
\caption{Medium-term effects and pre-COVID levels}
\includegraphics[width=\textwidth]{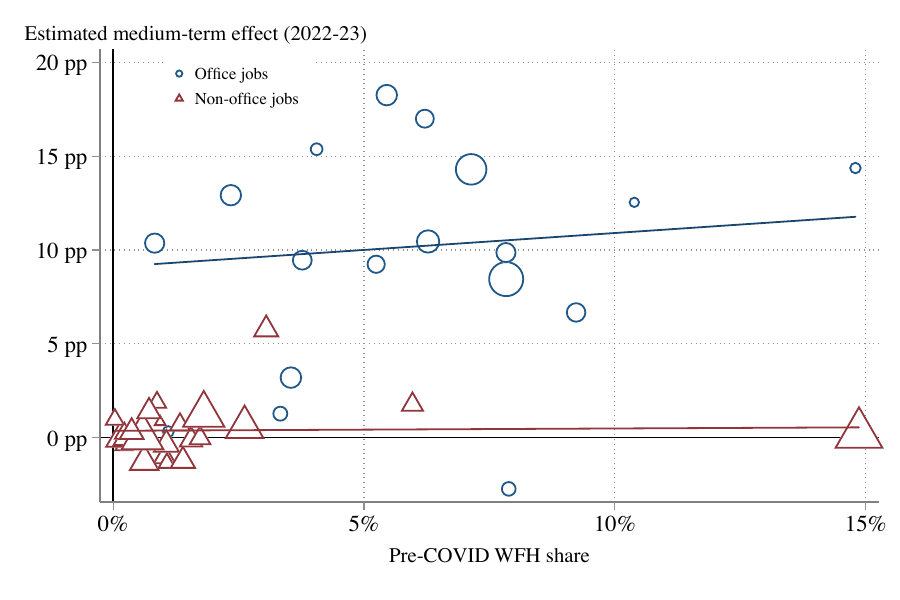} 
\end{subfigure}

\begin{minipage}{\textwidth}
{\footnotesize \underline{Notes}: These figures illustrate heterogeneity in WFH take-up across occupations during and after the pandemic in treated states, based on estimates of equation~\eqref{eq:basicDiD} with period-specific controls. Each point in the figure represents a major occupation group, with blue circles for office jobs and red triangles for non-office jobs. The size of the circle/triangle denotes the number of workers in the relevant occupation. The sample is restricted to workers in major urban areas.
\par} 
\end{minipage}
\end{figure}
\clearpage
\begin{figure}[htbp]
\centering
\caption{Correlation between short- and medium-term impacts on WFH share} \label{fig: correlation}

\includegraphics[width=\textwidth]{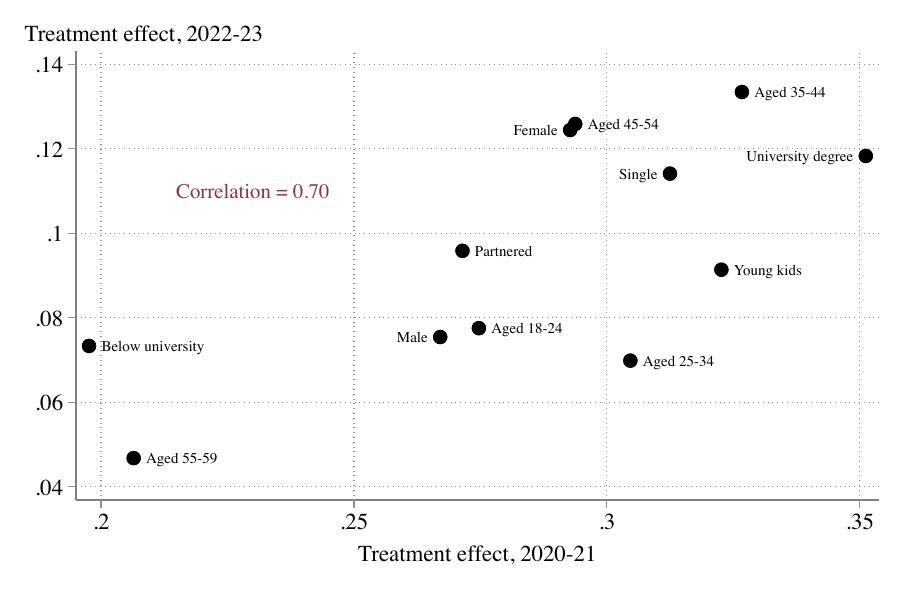} 

\begin{minipage}{\textwidth}
{\footnotesize \underline{Notes}: These figures illustrate heterogeneity in WFH take-up across subgroups of office workers in major urban areas during and after the pandemic. Each point represents the estimated treatment effect for a given subgroup.} 
\end{minipage}
\end{figure}


\clearpage 

\section{Summary of related studies and estimates}
\label{appendix_literature}

In Appendix Table~\ref{tab:summary_literature}, we benchmark our causal estimates of the persistent effects of lockdowns on WFH against related estimates in other correlational studies. Estimates come from two studies that involve the same six authors \citep{aksoy2022working, zarate2024does}. \cite{aksoy2022working} use data from the first two waves of the Global Survey of Working Arrangements (G-SWA), which surveyed full-time workers in 27 countries in the middle of 2021 and early 2022. \cite{zarate2024does} use data from wave 3 of G-SWA (2023, April--May), which involved workers from 34 countries, as well as 2023 respondents from the Survey of Working Arrangements and Attitudes (SWAA) in the United States. Both SWAA and G-SWA are online surveys initiated after the pandemic began. Both studies estimate the association between cumulative lockdown stringency (CLS) indices and WFH measures,\footnote{CLS is measured in each month is defined for a country/state as $\max$(shelter-in-place order, $0.75 \times $ business-closure order + $0.25 \times $  school-closure order).} and \cite{zarate2024does} include estimates using cross-country variation in CLS as well as variation within the US. Thus, there are three sets of estimates that are relevant to our analysis. For \cites{zarate2024does} estimates, the key outcome is the number of current WFH days per week. However, \cite{aksoy2022working} examine planned WFH days in 2022 and later, which we consider to be a more relevant variable for thinking about persistence than current WFH levels across 2021 and early 2022, given that vaccination coverage remained patchy and many countries still faced emergency restrictions at those times.

All three sets of estimates show the correlation between WFH measures and a one standard deviation increase in the CLS index. As a standard deviation varies across the three samples, we present the harmonized effect in Appendix Table~\ref{tab:summary_literature} of a one-month increase in lockdown. We harmonize our DiD point estimates in two steps. First, we convert the estimated increase in WFH share to full WFH days, assuming a five-day workweek, by multiplying the estimated increase in the WFH share (e.g., 0.100) by 5 (i.e., $0.1 \times 5$ = 0.5 workdays).\footnote{We calculate a similar effect if we use our estimates on WFH hours and assume an 8-hour workday.} Second, we divide these estimates by 4.7, which is the additional number of months of lockdown experienced by workers in treated states relative to control states on average.

We also try to get closer to the underlying samples used in other estimates. Our main sample consists of office workers in major urban areas~--- the main group targeted by remote work mandates. However, other studies do not restrict on occupation. Thus, we progressively broaden the sample to include non-office workers (comparable sample 1), urban workers (comparable sample 2) and all workers (comparable sample 3). We also show how our estimates vary when we restrict the sample to college graduates, as is done in other studies. Focusing on college graduates arguably makes the individuals in our sample more similar to international estimates, given the different job mix in different countries.

Appendix Table~\ref{tab:summary_literature} shows that our estimates are considerably larger than other estimates. For example, our estimates for college graduates in major urban areas (shown in bold) indicate that one more month of lockdown increases full WFH days by 0.082, compared to 0.021--0.028 in \cite{zarate2024does} and 0.036 for planned WFH days in \cite{aksoy2022working}.

Extending our sample to include minor urban and rural areas arguably makes our sample more comparable to other studies, but it comes at a cost. First, it makes the underlying parallel-trends assumptions less likely to be valid, as states become less comparable if we look beyond major urban areas. Second, lockdowns generally applied more strongly and for longer in major urban areas, which means that our harmonized estimates in comparable samples 2 and 3 will be underestimated. Nonetheless, even if we include all college graduates in the sample, our estimated harmonized effect of 0.053 is at least 50\% larger than other estimates.

\clearpage
\newgeometry{left=1.5cm, right=1.5cm, top=2cm, bottom=2cm}

\begin{landscape}

\tiny

\setlength{\tabcolsep}{2pt}
\begin{longtable}{>{\raggedright}p{3cm} >{\raggedright}p{5cm} >{\raggedright}p{4cm} >{\raggedright}p{4cm}>{\raggedright}p{3cm} >{\raggedright\arraybackslash}p{3cm}}
\caption{Summary of related estimates from other studies and comparison to our estimates} \label{tab:summary_literature} \\
\hline 
\textbf{Study} & \textbf{Setting, Data, Sample and outcomes} & \textbf{Measure: lockdown intensity} & \textbf{Empirical strategy} & \textbf{Coefficient estimate} & \textbf{Harmonized effect} \\ \hline 
\endfirsthead

\multicolumn{6}{c}{{\normalsize \tablename\ \thetable{}: Summary of related estimates from other studies and comparison to our estimates (continued from previous page)}} \\  
\multicolumn{6}{c}{} \\
\hline 
\textbf{Study} & \textbf{Setting, Data, Sample and outcomes} & \textbf{Measure: lockdown intensity} & \textbf{Empirical strategy} & \textbf{Coefficient estimate} & \textbf{Harmonized effect} \\  \hline 
 \endhead

 \hline \multicolumn{6}{r}{{Continued on next page}} \\ 
 \endfoot

\hline \hline
\endlastfoot
\addlinespace 

Aksoy et al. 2022. ``Working from home around the world.'' \textit{Brookings Papers on Economic Activity}, 2022(2): 281–360. & \underline{Setting}: 27 countries, mid 2021 and early 2022; \underline{Data}: Global Survey of Working Arrangements (G-SWA), waves 1 \& 2; \underline{Sample:} Age 20--59, full-time workers; finished primary school (N=~33,000 individuals); \underline{Outcomes:} current WFH days per week (as of mid 2021 and early 2022), planned WFH days in 2022 and later  & A 1 standard-deviation increase (4.7 months) in cumulative lockdown stringency at national level. Lockdown stringency in each month defined for country as $\max$(shelter-in-place order, $0.75 \times $ business-closure order + $0.25 \times $  school-closure order) & Individual-level correlational analysis with national measures of lockdown intensity (subnational in some cases); control variables for cumulative COVID deaths, wave and industry dummies, gender, education and age group & 4.7 months of lockdown stringency $\rightarrow$ 0.204 WFH days and 0.136 planned WFH days (Table 2, all respondents); 0.282 WFH days and 0.170 planned WFH days (Table 4, college graduates) & 
\underline{WFH days}: All: 0.043 (=0.204 days / 4.7 mths); \;\;\;\;\;\;\;\;\;\;\;\;\;\;\;\;\;\;\;\;       College: 0.060  \underline{Planned WFH days}: \; \;\;\;\;\;\;\;\;\;\;\;\;\;\;\;\;\;\;\;\; All: 0.028; \;\;\;\;\;\;\;\;\;\;\;\;\;\;\;\;\;\;\;\;                         \textbf{College: 0.036} \\  \hline  \addlinespace

Zarate et al. 2024 ``Why Does Working from Home Vary Across Countries and People?''. \textit{NBER WP No.\ 32374}. & \underline{Setting}: 34 countries, 2023 April-May; \underline{Data}: Global Survey of Working Arrangements (G-SWA), wave 3; \underline{Sample}: 20-64; full-time workers; secondary or tertiary education (N=34 country-level averages based on 42,426 individuals);  \underline{Outcome}: full paid days WFH per week & A 1 standard-deviation increase (6.2 months) in cumulative lockdown stringency at country level (2020-2022). Lockdown stringency in each month defined for country as $\max$(shelter-in-place order, $0.75 \times $ business-closure order + $0.25 \times $  school-closure order) & Cross-country correlational analysis with national measures of lockdown intensity; no pre period; four aggregate control variables (GDP per capita; weighted population density; individualism; industry mix) & 6.2 months of lockdown stringency $\rightarrow$ 0.08 WFH days (column 6, Table 3 - all respondents); 0.13 WFH days (column 6, Table 4 - college graduates). & 
All: 0.012 (= 0.08 days / 6.2 mths);  \; \;\;\;\;\;\;\; \;                        \textbf{College: 0.021}  \\ \hline \addlinespace

Zarate et al. 2024 ``Why Does Working from Home Vary Across Countries and People?''. \textit{NBER WP No.\ 32374}. & \underline{Setting}: US, 2023 April-May;  \underline{Data}: Survey of Working Arrangements and Attitudes (SWAA), 2023 respondent; \; \; \underline{Sample}: Age 20-64, worked 4+ days per week, earned \$10,000+ per year (N=$\sim$34,000 individuals);  \underline{Outcome}: full paid days WFH per week & A 1 standard-deviation increase (3.5 months) in cumulative lockdown stringency at state level (2020-2022). Lockdown stringency in each month defined for state as above.  &	Correlational analysis at the individual-level with state-level measures of lockdown intensity; no pre period; four aggregate control variables (average wage in state; population density of job ZIP code; county 2020 Joe Biden vote; industry WFH propensity) &	3.5 months of lockdown stringency $\rightarrow$ 0.06 WFH days (column 6, Table 5 - all respondents); 0.10 WFH days (column 6, Table 5 - college graduates). &	
All: 0.017 (= 0.06 days / 3.5 mths);  \; \; \; \; \; \;                          \textbf{College: 0.028}  \\ \hline \addlinespace

\underline{This paper}: Ketter et al. 2025. ``A new equilibrium: COVID-19 lockdowns and WFH persistence'' & \underline{Setting}: Australia; 2022 and 2023 (August-November); \underline{Data}: The Household, Income and Labour Dynamics in Australia (HILDA) Survey, 2002--23; \underline{Sample}; 18-59, employees (N=146,000ish observations);
\underline{Outcome}: WFH share of total working hours & Treatment: states that had, on average, 4.7 months more lockdowns in 2020-21 than other states & Differences-in-differences analysis comparing states with large differences in lockdown duration; long pre period and rich individual, job and location controls & 
\underline{Office workers, major} \underline{urban areas (main sample)}:
Treatment $\rightarrow$ WFH share 10 pp (Table 1), 11.8 pp (college) 
\underline{Major urban workers} \underline{(comparable sample 1)}:
Treatment $\rightarrow$ WFH share 5.1 pp (all), 7.7 pp (college)
\underline{Urban workers} \underline{(comparable sample 2)}:
Treatment $\rightarrow$ WFH share 3.8 pp (all), 6.9 pp (college) \; 
\underline{All workers} \underline{(comparable sample 3)}:
Treatment $\rightarrow$ WFH share 2.5 pp (all), 5.0 pp (college) 
& \underline{Main sample}: \; \;\; \;\; \;\; \;\; \;
All: 0.106; \; \;\; \;\; \;\; \;\; \;\; \;\; \;
College: 0.126

\underline{Comparable sample 1}: \; \;\; \;\; \;\; \;
All: 0.054;       \;\; \;\; \; \;\; \;\; \;\;\; \;\; \;                          \textbf{College: 0.082}
\underline{Comparable sample 2}: \; \;\; \;\; \;\; \;
All: 0.040; \; \;\; \;\; \;\; \;\; \;
College: 0.073 
\underline{Comparable sample 3}: \; \;\; \;\; \;\; \;
All: 0.026; \; \;\; \;\; \;\; \;\; \;
College: 0.053 \\
\end{longtable}
\end{landscape}


\restoregeometry

\end{appendices}
\renewcommand{\thepage}{A\arabic{page}}
\end{document}